\documentclass{caosp309}
\usepackage{graphicx}
\usepackage{natbib}
\articleNo{301}
\pubyear{2020}
\volume{50}
\volnumber{3}
\firstpage{1}
\received{May 1, 2020}
\accepted{July 28, 2020}

\def\BibTeX{{\rm B\kern-.05em{\sc i\kern-.025em b}\kern-.08em
             T\kern-.1667em\lower.7ex\hbox{E}\kern-.125emX}}

\begin{document}

\hauthor{U.\,Munari}

\title{Symbiotic Novae}

\author{
        Ulisse\,Munari\inst{1}\orcid{0000-0001-6805-9664}
       }

\institute{
           INAF National Institute of Astrophysics, Astronomical Observatory
           of Padova, 36012 Asiago (VI), Italy, \email{ulisse.munari@inaf.it}
          }

\date{October 8, 2024}

\maketitle

\begin{abstract}
(Invited Review)  According to modern definition, a symbiotic nova is an
otherwise normal nova (i.e.  powered by {\it explosive} thermonuclear
burning) that erupts within a symbiotic star, which is a binary where a WD
accretes from a cool giant companion.  Guided primarily by the very well
observed eruptions of RS Oph in 2006 and 2021, and that of V407 Cyg in 2010,
we describe the main multi-wavelength properties of symbiotic novae and
their relation to classical novae, and propose a 3D model structure that
identifies the emitting source location for hard and supersoft X-rays, radio
syncrothron and thermal, permitted and forbidden emission lines.  Very few
symbiotic novae are known in the Galaxy, and we compile a revised catalog
based on firm astrometric identification.  The exciting prospect of an
imminent new outburst of T CrB is also discussed.
\end{abstract}

\section{A modern definition for symbiotic novae}
\label{intr}

Traditionally \citep[eg.][]{1957gano.book.....G, 1980MNRAS.192..521A,
1986syst.book.....K, 1995cvs..book.....W}, the term 'symbiotic nova' has
indicated a symbiotic star undergoing a large amplitude ($\Delta V$$\geq$ 4
mag) and very long lasting outburst (several decades to a century), clearly
distinct from the smaller amplitude ($\Delta V$$\sim$2 mag) and much faster
eruptions (from several months to a few years) characterizing most of the
symbiotic binaries and usually referred to as 'ZAND' events, from the
prototype Z And that has gone through several of them in its recorded
history \citep[eg.][]{2016MNRAS.462.4435T, 2020A&A...636A..77S}. Over the
last 10-20 yrs the accepted meaning has however drastically changed to
\begin{itemize}
\item[] {\it a symbiotic nova is an otherwise normal nova [i.e.  powered
by explosive thermonuclear burning] that erupts within a symbiotic binary}
\end{itemize}
\noindent
which we will adopt in this review.  Classical novae and symbiotic novae
share similar properties in the way their white dwarfs (WD) accrete mass and
then ignite the explosive burning.  A classical nova is a compact
interacting binary, where a WD accretes (usually via Roche lobe overflow and
formation of an accretion disk) from a low-mass red dwarf companion (RD),
orbiting in a few hours at a distance of about a Solar radius.  In the case
of a symbiotic binary the companion to the WD is a red giant (RG), and to
accommodate its much larger dimensions the orbital separation widens to a
few astronomical units and the orbital period increases to a few years.

For both the classical and symbiotic novae, the accreted material goes
accumulating in electron-degenerate conditions on the surface of the WD;
diffusion from the outer layers of the WD allows enrichment in CNO nuclei
(ensuing stronger explosions).  Once the pressure at the base of the
accumulated shell reaches a critical value around 10$^{20}$ dyn/cm$^{2}$,
nuclear burning of the accreted material sets in.  Given its
electron-degenerate state ($P\propto \rho^\beta$), the shell does not
initially react by expanding to the increase in temperature, with the
consequent ramping in energy released by the nuclear reactions
($\epsilon_{\rm CNO} \propto$ T$^{17}$) which leads to further increase in
temperature.  Within a few minutes, the temperature reaches the Fermi value
(of the order of 10$^8$~K), at which point the electron degeneracy is
suddenly lifted ($P\propto \rho T$), and the hyper-hot material in the shell
violently expands at velocities in excess of the escape velocity from the WD
($v_e=\sqrt{2GM/R}\sim 9100$ km\,sec$^{-1}$ at $M$=1.25\,M$_\odot$;
\citet{2024ApJ...963...17A}), effectively terminating the TNR
\citep[eg.][]{1989clno.conf...39S,2008clno.book...77S}.  Not the
entire shell is however expelled, and the nuclear burning can continue (now
in thermal equilibrium) in what has been left of the shell on the surface of
the WD, a phase lasting for weeks--months and characterized by supersoft
X-rays emission, which become observable
when the ejecta turn optically-thin \citep[eg.][]{2011ApJS..197...31S}.

The paths followed by classical and symbiotic novae diverge at this point,
at the time when the shell is violently expelled by the WD.  Unbind from the
system, the ejected material travels the WD--RD orbital separation of a
classical nova in a matter of minutes, after which (and for the time being)
it basically keeps expanding on free ballistic trajectories into the
circumstellar void, as beautifully demonstrated by Nova Per 1901
\citep{2012ApJ...761...34L} and other novae with spatially resolved ejecta
\citep{2008clno.book..285O}.  In a symbiotic nova instead, the circumstellar
space up to hundreds of AU is filled by the thick and slow wind of the RG,
and the fast expanding ejecta have to plow their way through it
\citep{2008ApJ...685L.137S, 2008ASPC..401.....E}.  This results in a violent
and efficient deceleration of the ejecta, and the associated massive
shock-fronts power emission from the highest energies probed by Cherenkov
telescopes all the way down to the synchrotron emission dominating at radio
wavelengths.  The first nova ever detected in $\gamma$-rays (by the {\it
Fermi}-LAT satellite) has been, not incidentally, the symbiotic nova V407
Cyg in 2010 \citep{2010Sci...329..817A}.

The literature about observations and modeling of nova explosions is very
rich, and well beyond the possibility of a balanced citation in a short
review like the present one.  Excellent summaries may be found in the books
and proceedings of conferences devoted specifically to novae, like 
those edited by \citet{1989clno.conf.....B}, \citet{1990LNP...369.....C},
\citet{2002AIPC..637.....H}, \citet{2008clno.book.....B},
\citet{2012clno.book.....S}, and \citet{2014ASPC..490.....W}.

\underline{\sc Old meaning for 'symbiotic novae'}.  Before moving on, some
words are in order about the {\em traditional} meaning of the term
'symbiotic nova', which is still - although rarely - in use.  In the
life-cycle of a typical symbiotic star as proposed by
\citet{2019arXiv190901389M}, the longest intervals are spent accreting onto
the WD (at highly variable rates).  In the (great) majority of the symbiotic
stars the accreted envelope is non-degenerate, primarily because of the low
mass of the WD.  Once the conditions for nuclear burning are reached, the
burning ignites non-explosively and continues in thermal equilibrium as long
as enough hydrogen fuel is present in the shell, a phase lasting decades to
centuries \citep{1982ApJ...259..244I, 1982ApJ...257..752F,
1982ApJ...257..767F,1983ApJ...273..280K}.  The rise to optical maximum takes
a few years to complete ($\sim$3 yrs for V1016 Cyg and V3268 Sgr): these are
the symbiotic novae according to the old definition (for which we propose
the name SETE standing for Symbiotic stars Erupting in Thermal-Equilibrium). 
The shell is {\it not} ejected and remains in hydrostatic equilibrium,
progressively contracting in radius and slowly rising in surface temperature
as the hydrogen fuel is consumed.  Once the burning shell shrinks to the
radius and temperature of the nuclei of planetary nebulae, accretion-induced
ingestion by the shell of limited amounts of mass triggers short-lived ZAND
outbursts \citep[as observed in 2015 for AG Peg;][]{2016MNRAS.462.4435T}. 
The exhaustion of the hydrogen fuel in the shell eventually stops the
burning, the strong emission lines and ultraviolet continuum die away, and
the symbiotic star (now broadly resembling a field single RG) resumes the
long-lasting and quiet accretion phase in preparation for a new cycle.

\section{A revised catalog of Symbiotic Novae}

From the previous section, the main difference between classical and
symbiotic novae resides in the nature of the companion to the WD: a red
dwarf versus a red giant. Considering that in the infrared $K$ band
($i$) cool dwarfs and cool giants differ by $\sim$10 in absolute M(K)
magnitude, and ($ii$) the interstellar reddening plays a marginal role
(A(K)=0.43$\times$E$_{B-V}$ for M spectral types), and ($iii$) the
sensitivity of the 2MASS $J$$H$$K$ survey is high enough to detect a red
giant anywhere in the Galaxy, it seems a good choice to investigate the
distribution of novae in 2MASS $K$ band to segregate symbiotic novae from
classical ones.

We started by assembling various compilations of Galactic Novae (from IAU,
SIMBAD, Bill Gray, Koji Mukai, and others) and proceeded with pruning them
from mis-entries.  For ex., FU~Ori - the prototype of erupting pre-main
sequence stars - is still listed as Nova Ori 1939.  We therefore retained
only the objects for which there is observational evidence of an {\it
explosive} TNR event on a WD (which excludes SETE objects).  Next, we tried
to obtain accurate astrometric coordinates and a measure of their error by
going to the original sources of the epoch (for ex., SIMBAD coordinates for
Nova Sgr 1935 are those of an unrelated Mira variable, $\sim$1 arcmin away
from the true nova position, as clear from a finding chart of the epoch). 
For novae up to 1985, the prime source for astrometric coordinates and
finding charts is the reference catalogue and atlas of
\citet{1987rcag.book.....D}, where in most cases the original discovery
photographic plates were directly inspected and astrometrically measured in
a modern way.  For later novae, we went through the original discovery
announcements (on IAUC, CBET, ATel, etc.), follow-up papers, and the catalog
and atlas of CVs by \citet{1993PASP..105..127D} and \citet{1997PASP..109..345D,
2001PASP..113..764D}.  The quoted coordinates were critically re-evaluated
and their errors estimated in parallel with direct inspection of POSS scans
and 2MASS maps.  Only novae identified with a high degree of confidence and
not in blend with field stars were retained.  The surviving objects were
then cross-matched with Gaia DR3 \citep{2016A&A...595A...1G,
2023A&A...674A...1G} looking for possible nearby, faint field stars that
could interfere with a firm identification of the progenitor.  After this
further pruning, the Gaia DR3 coordinates were adopted and the final list of
novae was cross-matched with the 2MASS catalog \citep{2006AJ....131.1163S}
within 1arcsec radius (including correction for proper motion).

\begin{figure}
\centerline{\includegraphics[width=0.75\textwidth,clip=]{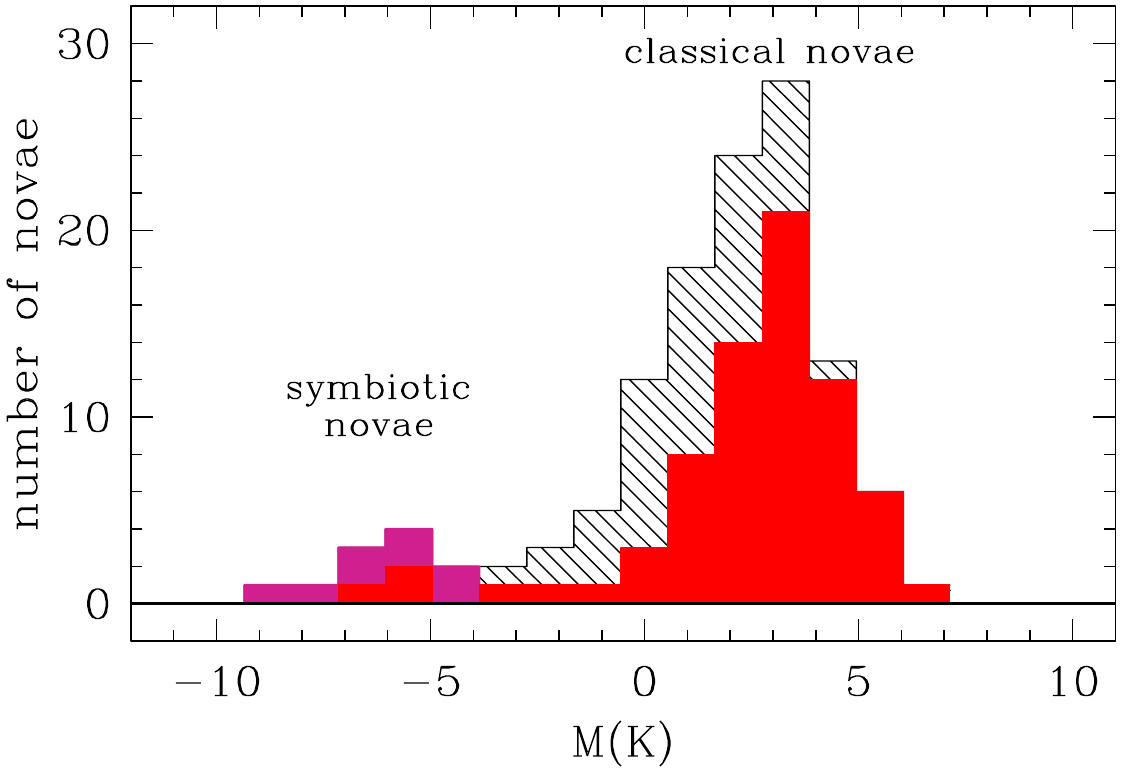}}
\caption{Distribution in absolute M($K$) magnitude of all validated 
thermonuclear runaway novae with an accurate astrometric identification.
Those containing a red giant (M($K$)$\leq$$-$4 mag) are listed in
Tab.\,\ref{tabSyN}.}
\label{figSyN}
\end{figure}

\begin{table}[t]
\small
\begin{center}
\caption{The symbiotic novae (novae with a red giant), corresponding to the
objects with M($K$)$\leq$$-$4 in Fig.\,\ref{figSyN}.}
\label{tabSyN}
\centerline{\includegraphics[width=\textwidth,clip=]{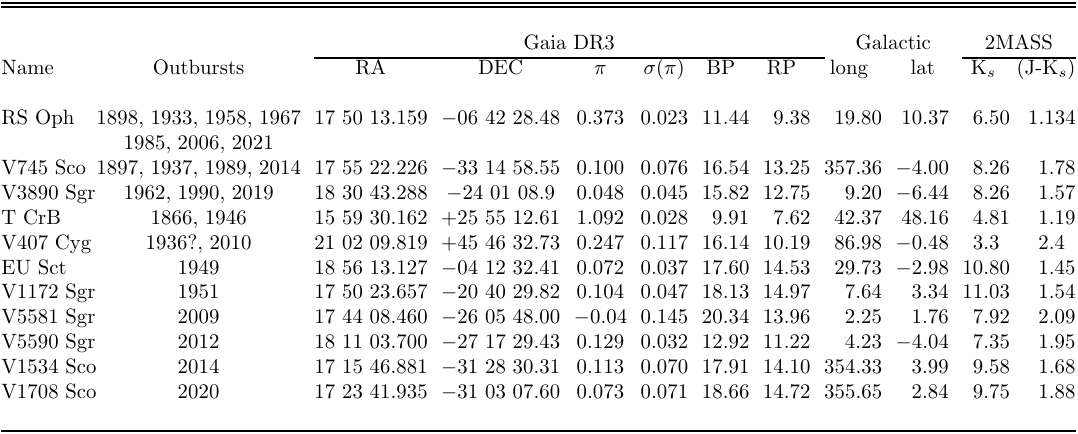}}
\end{center}
\end{table}

As a final step, the Gaia DR3 parallax was adopted to compute the absolute
magnitude of novae in the $K$ band (2.2 $\mu$m), which resulted in the
distribution of Fig.~\ref{figSyN}.  In this figure, the novae with an error
on the parallax inferior to 25\%, are plotted in red, those with a larger
error are dashed.  Various novae with a red giant are listed in Gaia DR3
with a too large uncertain in the parallax, and consequently we derived
their distances as an average of independent estimates found in literature;
these systems are plotted in purple in Fig.~\ref{figSyN}.  No correction for
reddening is applied because it is reliably {\it measured} (not simply
guessed) only for a tiny minority of all novae, and its impact - which is at
the level of a few tenths of a magnitude at most - does not affect at all
the distinction between classical and symbiotic novae in Fig.~\ref{figSyN}.

The distribution in Fig.~\ref{figSyN} is clearly bi-modal.  The more
numerous group, which is that of classical novae (those with a red dwarf),
presents a Gaussian-like distribution centered at M($K$)=+3.2 mag.  It is
worth noticing that this value is about 1 mag brighter than the absolute $K$
magnitude of main sequence K-type stars, suggesting that the same accretion
disk that dominates the UV and optical energy distribution of old novae
\citep{1990apcb.conf..373C, 2004ESASP1283.....C} extends its influence also
into the infrared.

The smaller group in Fig.~\ref{figSyN}, is that of symbiotic novae, with a
mean absolute magnitude of M($K$)$\approx$ $-$5.8 mag, which is close to the
mean M($K$)= $-$5.9 mag derived by \citet{2021MNRAS.505.6121M} for the
accreting-only symbiotic stars.  This similarity can be read as an
indication that the population of red giants in field symbiotic stars and in
symbiotic novae is the same, and presumably also their mass lose rates. 
The symbiotic novae in Fig.~\ref{figSyN} (those with $M(K)\leq -4.0$ mag)
are listed in Tab.~\ref{tabSyN}, with quoted values taken from Gaia DR3 and
2MASS catalogues.  Those with more than one recorded eruption (symbiotic
{\it recurrent} novae) are listed at the top.  

\section{The distinctive presence of the red giant wind}

We remarked above how the primary differences between classical and
symbiotic novae originate from the wind of the red giant filling the
circumstellar space of symbiotic novae, and how it interacts with the
radiation output and the fast ejecta of the outbursting WD.

\subsection{Flash ionization of the red giant wind}

A feature unique to symbiotic novae actually manifests even before the fast
ejecta begins slamming onto the pre-existing slow wind of the RG.  It
originates from the strong UV-flash originating from the burning shell
during the TNR and its immediate aftermath.  The flash is really short-lived:
in the nova models by \citet{2008clno.book...77S}, the shell surface
temperature drops to 1$\times$10$^5$~K in $\sim$300sec and $\sim$500sec
for WDs of 1.35 and 1.25~M$_\odot$ mass, respectively. 

In a classical nova, such a short-lived UV-flash precedes the discovery of
the nova at optical wavelengths by $\sim$one day, and will go unnoticed
while traveling through the empty circumstellar space.  On the contrary, in
a symbiotic nova, the RG wind that permeates the circumstellar space absorbs
the photons of the UV-flash and get ionized.  The ionized gas starts
immediately to recombine, glowing very brightly in the process.  The
recombination time scale (in hours) goes as $t_{\rm rec} = 0.66\,
T_{e}^{0.8}\, n_{e}^{-1}$ \citep{2003ARA&A..41..517F}, where $T_e$ is the
electron temperature in 10$^4$~K units, and $n_e$ is the electron density in
10$^9$~cm$^{-3}$ units.  Both lines and continuum are emitted by the
recombining wind.

\begin{figure}
\centerline{\includegraphics[width=\textwidth,clip=]{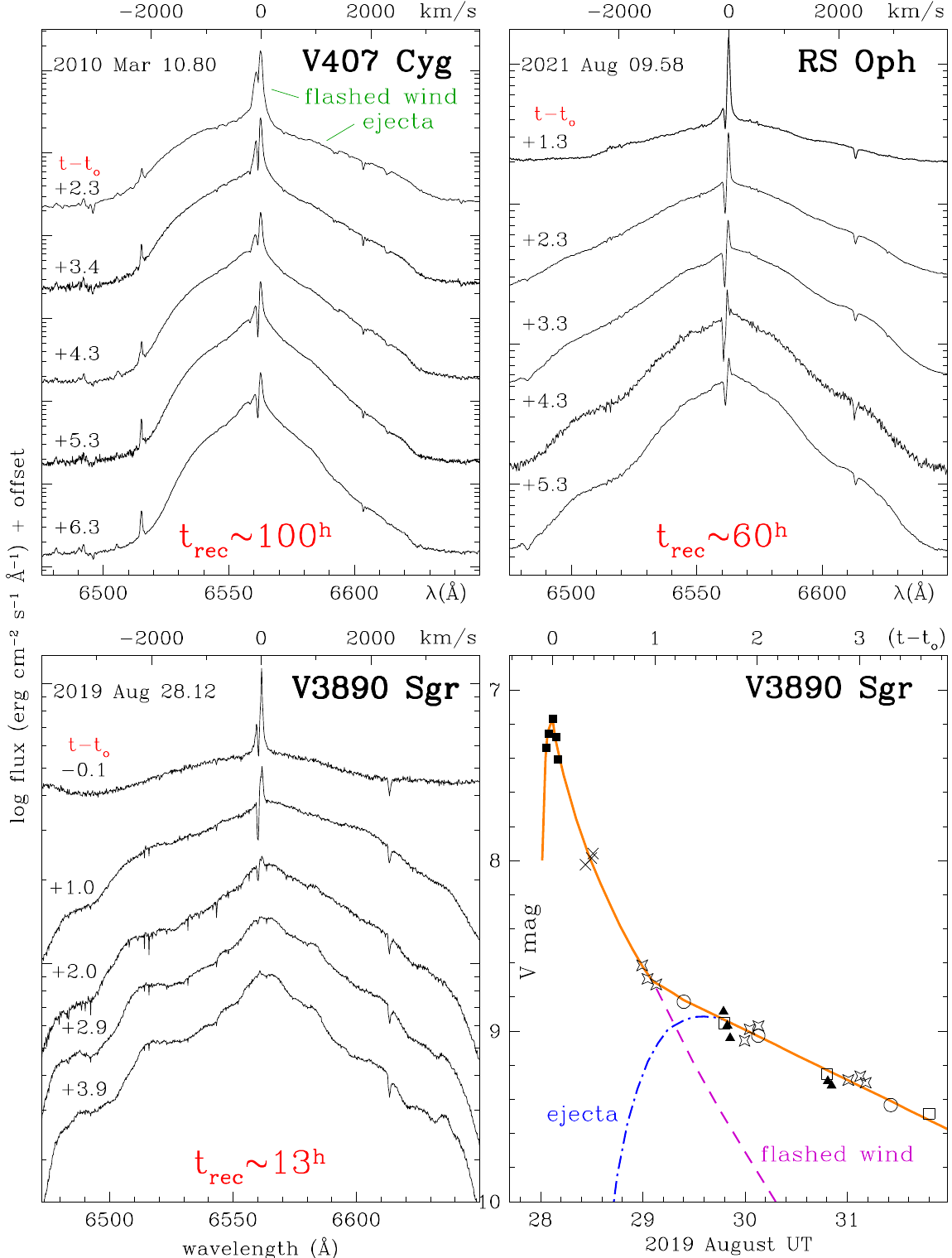}}
\caption{High-resolution H$\alpha$ profiles covering the first days
for the outburst of three symbiotic novae, to highlight the sharp
component from the recombining flashed-wind of the RG and the broad one from
the expanding ejecta. For comparison, the early photometric evolution of 
V3890 Sgr is deconvolved into the same two components \citep[data 
from][F.~Walter private~comm.]{2011MNRAS.410L..52M, 2021arXiv210901101M}}
\label{figflash}
\end{figure}

\underline{\sc Emission Lines}.  The emission lines component from the
flashed wind are quite sharp (FWHM$\sim$40 km\,sec$^{-1}$, as the outflow
velocity of the RG wind is rather low), and are visible on-top the broad
pedestal originating from the fast expanding ejecta (see  top-left of
Fig.~\ref{figflash}).  The evolution of the recombining H$\alpha$ component
over the first few days of the V407~Cyg (2010), RS Oph (2021), and V3890 Sgr
(2019) outbursts is compared in Fig.~\ref{figflash}.  The e-folding
recombination times are respectively 100, 60 and 13 hours, corresponding to
electronic densities of 6.6$\times$10$^6$, 1.1$\times$10$^7$ and
5.0$\times$10$^7$\,cm$^{-3}$.  From Fig.~\ref{figflash} its is clear that
prompt access to a telescope with a standing-by high-resolution spectrograph
(resolving power $\geq$20,000) is mandatory to successfully detect the
presence of emission lines from the recombining wind and record the details
of their very fast evolution.

The narrow emission lines from the flashed wind are invariably accompanied
by a similarly sharp and slightly blue-shifted {\it absorption} component. 
It originates from the outer parts of the slowly outflowing RG wind, which
remains neutral through the recombination process (which is equivalent to
say that there is enough column density in the circumstellar material to
absorb {\it all} the ionizing photons emitted during the initial UV-flash). 
While the narrow emission component traces gas which is roughly symmetric
around the central binary, the blue-shifted narrow absorption provides a
measure of the terminal wind velocity (at least in the direction probed by
the line-of-sight to the observer).  From the spectra in
Fig.~\ref{figflash}, such terminal wind velocity is $-$38, $-$60, and $-$65
km\,sec$^{-1}$ for V407~Cyg, RS Oph, and V3890 Sgr, respectively.

\underline{\sc Nebular continuum}.  If the passage of a classical nova at
peak optical brightness corresponds to the maximum angular extension of the
optically-thick pseudo-photosphere of the expanding ejecta, in a symbiotic
nova it generally marks the time when the continuum radiation from the
recombining wind reaches its peak output, roughly corresponding to the
maximum mass reached by the ionized fraction of the RG wind.  For ex.,
in the case of the 2019 outburst of V3890 Sgr, the continuum emission from
flashed wind reached a peak brightness about 4$\times$ that of the ejecta,
as illustrated by the deconvolution of the $V$-band lightcurve in
Fig.~\ref{figflash}.  The recombination time of the flashed wind is,
unsurprisingly, the same if estimated from the continuum emission or from
the narrow H$\alpha$ component, as illustrated for V3890 Sgr in
Fig.~\ref{figflash}, providing in both cases an $e$-folding time of $\sim$13
hours.

\underline{\sc Radio thermal emission}.  The UV-flashed wind of the RG emits
free-free radiation that should be detectable at radio wavelengths within
1-2 days of TNR onset, before the stronger synchrotron emission from the
shocked ejecta takes over.  The geometrical arrangement and flux level for
this prompt free-free radio emission should be similar to that of the RG
wind kept ionized by the long-term burning WDs in SETE objects, which has
been modeled by \citet{1984ApJ...286..263T}, among others.  A fair fraction
of symbiotic stars with burning WDs is radio-emitter at cm-wavelengths, with
brightness temperature and $\alpha$ spectral index ($S_\nu \propto
\nu^\alpha$) suggestive of thermal emission from an outflowing wind
\citep[eg.][]{1984ApJ...284..202S, 1990ApJ...349..313S,
1993ApJ...410..260S}.  The main difficulty to observe at radio wavelengths
this prompt free-free emission resides in the (very) fast response time
required to access suitable radio telescopes before the recombination wave
(proceeding inside-out because of the density gradient in the RG wind)
completes its job and the synchrotron emission from the shocked ejecta takes
over.

The earliest radio observation for the 2014 outburst of V745 Sco started 2.6
days past optical maximum \citep{2024MNRAS.534.1227M}, 2.0 days for V3890
Sgr in 2019 \citep{2023MNRAS.523.1661N}, and 2 to 5 days for RS Oph in 2021
\citep{2021ATel14849....1W, 2021ATel14886....1S}.  They all were already
affected by the rapidly growing contribution of the synchrotron emission
from the shocked ejecta, at an epoch when the thermal emission from the
flashed wind was declining if not already extinguished.  Better luck could
favor the coming and much anticipated outburst of T CrB (in 2025-2026 ?),
for which proposals have been approved at basically all radio facilities to be
triggered by the announcement of a new outburst. A few of them
specifically aim to obtain a "picture at rest" of the circumstellar medium
before the ejecta will slam onto it, by imaging the prompt free-free thermal
emission from the UV-flashed wind within hours of the outburst discovery.

\subsection{Deceleration of the ejecta by the red giant wind} 

The second feature unique to symbiotic novae is the violent deceleration of
the nova ejecta as they plow through the pre-existing circumstellar
material, fed by the slow wind of the RG.  The ejecta are shocked and give
origin to copious non-thermal emission over the entire wavelength range,
from radio synchrotron to the TeV energies mapped by Cherenkov telescopes
\citep{2022NatAs...6..689A, 2022Sci...376...77H}.  For RS Oph, the
multi-wavelength results gathered from the 2006 and 2021 eruptions are very
similar, which may be read as an indication that the cavity shaped by
preceding eruptions is refilled by the wind of the red giant well in time
for the following outburst, a results probably applicable also to the other
symbiotic novae that were less intensively observed compared to RS Oph.

\underline{\sc DEOP}.  The multi-wavelength appearance and fate of the
shocked ejecta rest ultimately with the geometric arrangement of the
circumstellar medium.  Without the presence of the WD companion, the wind of
the RG would fill the circumstellar space in a roughly spherically-symmetric
manner.  The gravitational pull of the WD deflects instead toward the
orbital plane the mass-loss from the RG
\citep[eg.][]{2012BaltA..21...88M,2015A&A...573A...8S}, creating a {\it
Density Enhancement on the Orbital Plane} (DEOP).  The more wind is
deflected toward DEOP, the less is left to flow perpendicular to the
orbital plane: ejecta launched into polar directions then suffer from only
limited deceleration, while those on the orbital plane slam onto a
much denser medium able to efficiently absorb (all) their kinetic momentum.
For a given WD mass and RG mass-loss rate, a WD orbiting closer to the 
RG is more efficient in deflecting the RG wind away from polar directions 
and onto DEOP.

\begin{figure}
\centerline{\includegraphics[width=\textwidth,clip=]{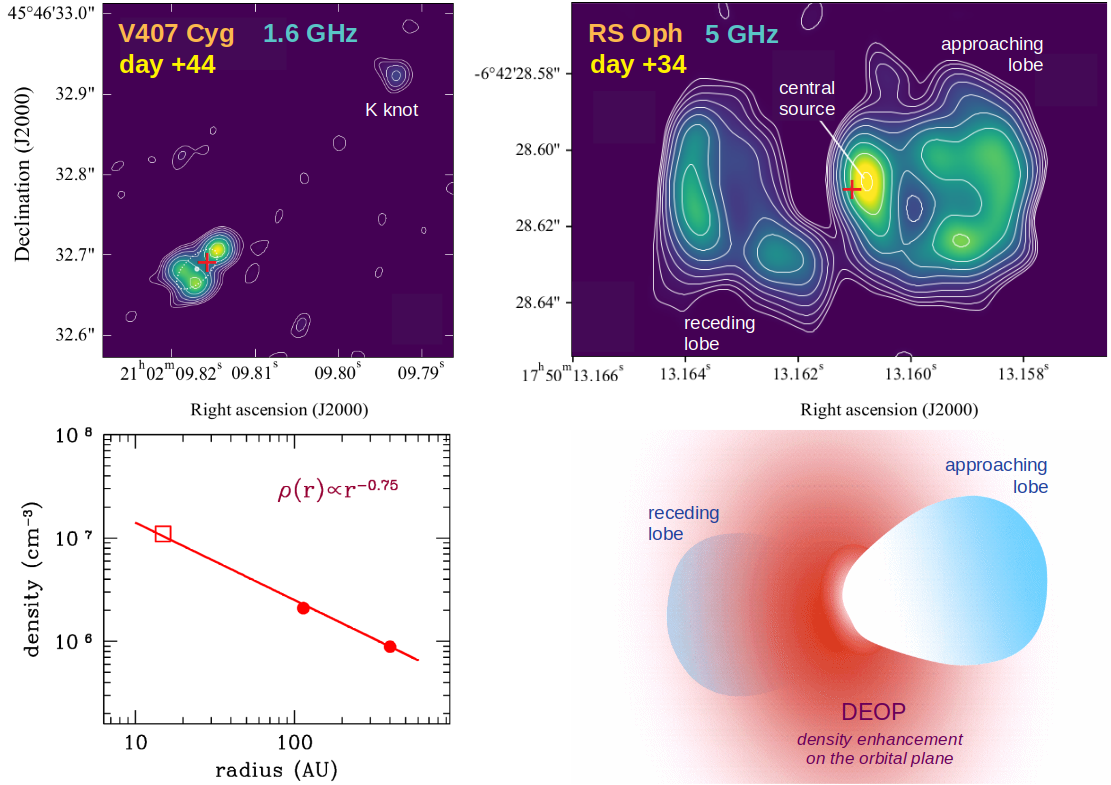}}
\caption{Top: VLBI high resolution radio maps of the 2010 and 2021 outbursts
of the symbiotic novae V407~Cyg and RS~Oph.  The red cross marks the
astrometric position from Gaia.  Bottom right: cartoon highlighting the
location of DEOP and the extinction it excerpts on the receding lobe of RS
Oph (compare with panel directly above).  Bottom left: the radial dependence
of density within DEOP of RS Oph \citep[data from][]{2020A&A...638A.130G,
2022A&A...666L...6M, 2024A&A...692A.107L}.}
\label{figEVN}
\end{figure}

\underline{\sc Bipolar radio-synchrotron lobes}.  Direct imaging of the
expanding eje-cta of RS Oph, years past the 2006 outburst, have been obtained
with HST in [OIII]~5007 line \citep{2009ApJ...703.1955R} and with Chandra in
X-rays \citep{2022ApJ...926..100M}.  Detecting such faint signals from
years-old ejecta requires careful PSF-subtraction and image deconvolution. 
More direct and detailed images of the expanding ejecta can be obtained in
the radio with VLBI techniques (very long baseline interferometry).  Their
milli-arcsec capabilities allow prompt resolution of the structure of the
ejecta already within a few weeks of the eruption.  An example of the
results obtained with EVN (European VLBI Network) for the outbursts of V407
Cyg (2010) and RS Oph (2021) are shown in Fig.~\ref{figEVN}: the
radio-emitting ejecta are shaped into wide bipolar lobes expanding
perpendicularly to DEOP.  The radio emission is of synchrotron origin, as
supported by the absorption-corrected $\alpha$ spectral slope and by the
huge brightness temperature (10$^7$/10$^8$~K), which is orders of magnitude
hotter than associated to thermal processes (10$^4$~K).

\begin{figure}
\centerline{\includegraphics[angle=270,width=\textwidth,clip=]{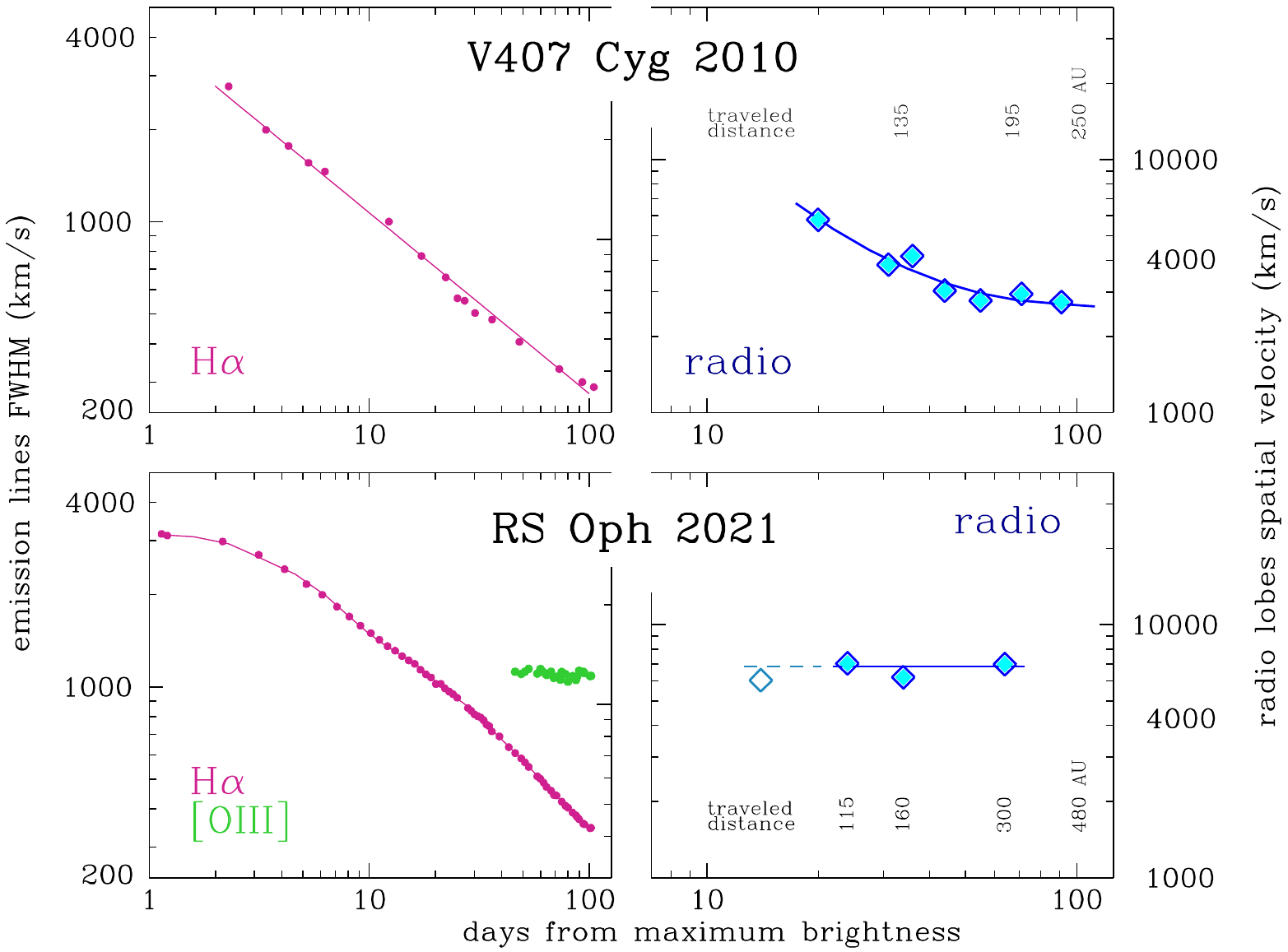}}
\caption{Comparing the evolution of the FWHM of Balmer and [OIII] emission
lines (left) and the space velocity of the radio bipolar jets (right) for
V407 Cyg and RS Oph \citep[data from][]{2011MNRAS.410L..52M,
2020A&A...638A.130G, 2022arXiv220301378M, 2024A&A...692A.107L}.}
\label{figfwhm}
\end{figure}

The angular expansion of radio lobes can be transformed into corresponding
space velocity by knowledge of the distance and the orbital inclination,
fairly well known for both V407 Cyg and RS Oph (4.0 and 2.7 kpc,
$\sim$80$^\circ$ and 57$^\circ$\footnote{The orbital inclination of RS Oph
is here revised from $i$=54$^\circ$ derived by \citet{2022A&A...666L...6M}
to $i$=57$^\circ$ by considering the improved radio angular expansion
rates from \citet{2024A&A...692A.107L}.}, respectively).  The time dependence of the
space velocity of their radio lobes is compared in Fig.~\ref{figfwhm}.  The
radio lobes of RS Oph have kept expanding at a high and constant
$v_{exp}$=8150~km\,s$^{-1}$ space velocity since the earliest radio epoch
(day +14), which indicates that ($i$) the deceleration along the polar
directions was completed already within the first two weeks, ($ii$) the WD orbiting
close to the RG \citep[1.5 AU orbital separation,][]{2009A&A...497..815B}
has been very effective in deflecting toward DEOP the wind of the RG, with a
minimal amount of material present above 50~AU on the polar directions, and
($iii$) the terminal space velocity of the radio lobes is still a fairly
large fraction of the initial launch velocity.  The radio lobes of V407~Cyg
have been instead decelerating all the way through the radio observing epochs,
slowing from 6000~km\,s$^{-1}$ on day +20 to 2800~km\,s$^{-1}$ on day +91. 
This leads to conclude that the WD orbiting the RG at great distance
\citep[$>$ 30 AU orbital separation,][]{2013ApJ...770...28H} was not
particularly effective in deflecting the wind of the RG away from polar
directions, with enough gas still present 250~AU above DEOP to keep the
ejecta of V407 Cyg decelerating for months.  Considering the velocities
observed in RS Oph and V407 Cyg, and the respective deceleration profiles, it
seems safe to infer that the initial launch velocity was similar for them
both and probably close to 10,000/12,000~km\,s$^{-1}$.

The radio lobes of RS Oph, thanks to the 57$^\circ$ orbital inclination,
have allowed to directly {\em "see"} DEOP for the first time in a symbiotic
system.  DEOP in RS~Oph has been kept ionized by the WD burning nuclearly
for the first three months of the outburst, so covering the whole set of
radio observing epochs (day +14 to +64).  An ionized DEOP is very effective
in producing free-free absorption of the synchrotron radiation from the
receding lobe which moves in the background to DEOP.  Because the density
within DEOP declines radially away from the central binary (cf.  the
lower-left panel of Fig.~\ref{figEVN}), also the free-free opacity declines
in pace, with the net result that the receding lobe becomes visible only
once its leading edge has moved behind DEOP outer regions of sufficiently
low optical depth, which happens around day +20/+21, following the
geometrical arrangement sketched in the lower-right panel of
Fig.~\ref{figEVN}.

Before leaving this section about large-scale radio structures, it is worth
mentioning an interesting and unexpected feature visible in the radio map of
Fig.~\ref{figEVN} for V407 Cyg: a knot of emission (labelled "K"), appears
moving radially away - and along the same direction as the bipolar lobes -
at a velocity (700~km\,sec$^{-1}$) that projects it back to the central
binary $\sim$6.5 yrs before the 2010 eruption, when V407 Cyg was undergoing
a spectacular surge in the accretion rate onto the WD
\citep{2003ARep...47..777K}, similar to what happened to T CrB in the 8 yrs
preceding its 1946 outburst \citep{1946ApJ...104...75P}.  Clearly, during
such super-active accretion phases (SAP) preceding the nova explosion, the
WD of symbiotic novae can expel isolated and massive blobs of material along
the orbital polar axes.  It will be worth searching for similar blobs on
radio maps collected at the time of the next eruption of T CrB, considering
the new SAP phase it has recently undergone during 2015-2023
\citep{2016NewA...47....7M}.  It will also be worth to re-inspect the abundant
spectroscopic material collected during the SAP phase of V407 Cyg
\citep[eg.][]{2003ARep...47..889T} and search for spectral counterparts to
the ejection of the "K" radio blob.

\underline{\sc Shrinking of emission lines}.  An easily observable
manifestation of the violent deceleration of the ejecta in a symbiotic nova
is the rapid narrowing of the {\it broad component} (cf.  identification at
top-left of Fig.~\ref{figflash}) of the emission lines
\citep[eg.][]{2011A&A...527A..98S, 2011MNRAS.410L..52M,
2014ApJ...785L..11B}.  The left panels of Fig.~\ref{figfwhm} show the
shrinking of H$\alpha$ for the recent outbursts of V407 Cyg and RS~Oph,
characterized by slopes (FWHM\,$\propto$\,$t^{\theta}$) of $\theta$=$-$0.59
and $-$0.73, respectively [for simplicity, in the rest of this paper
H$\alpha$ will be used as proxy for permitted emission lines in general]. 
The narrowing of H$\alpha$ proceeds very smoothly, supporting an origin in
ejecta that expand through a medium that extends also very smoothly over the
traveled distance, without discontinuities or sudden changes in the radial
density gradient.  By integrating over the behavior in Fig.~\ref{figfwhm},
and taking $v_{exp}$=FWHM/2.355 (i.e.  the $\sigma$ of the Gaussian
fitting) as representative of the bulk velocity of the gas, the distance
traveled by the H$\alpha$ emitting ejecta in the first 100 days is 15 and 14
AU for V407 Cyg and RS Oph, respectively.  Such a limited dimension fits
very nicely with the central synchrotron radio source (cf upper-right of
Figure~\ref{figEVN}) of RS Oph, that when visible for the first 40 days past
the initial outburst remained unresolved at the EVN restoring ellipsoidal
beam of 10$\times$12 milli-arcsec, corresponding to an upper limit of 13 AU
to its radius.  The fact that over the initial 100 days the radio lobes
travel for 250-500 AU in the polar directions, while the H$\alpha$ emitting
ejecta cover just 14-15 AU under constant deceleration, clearly indicates
that the (main) source of H$\alpha$ emission is the ejecta moving close to the
orbital plane and slamming onto DEOP.

\begin{figure}
\centerline{\includegraphics[width=\textwidth,clip=]{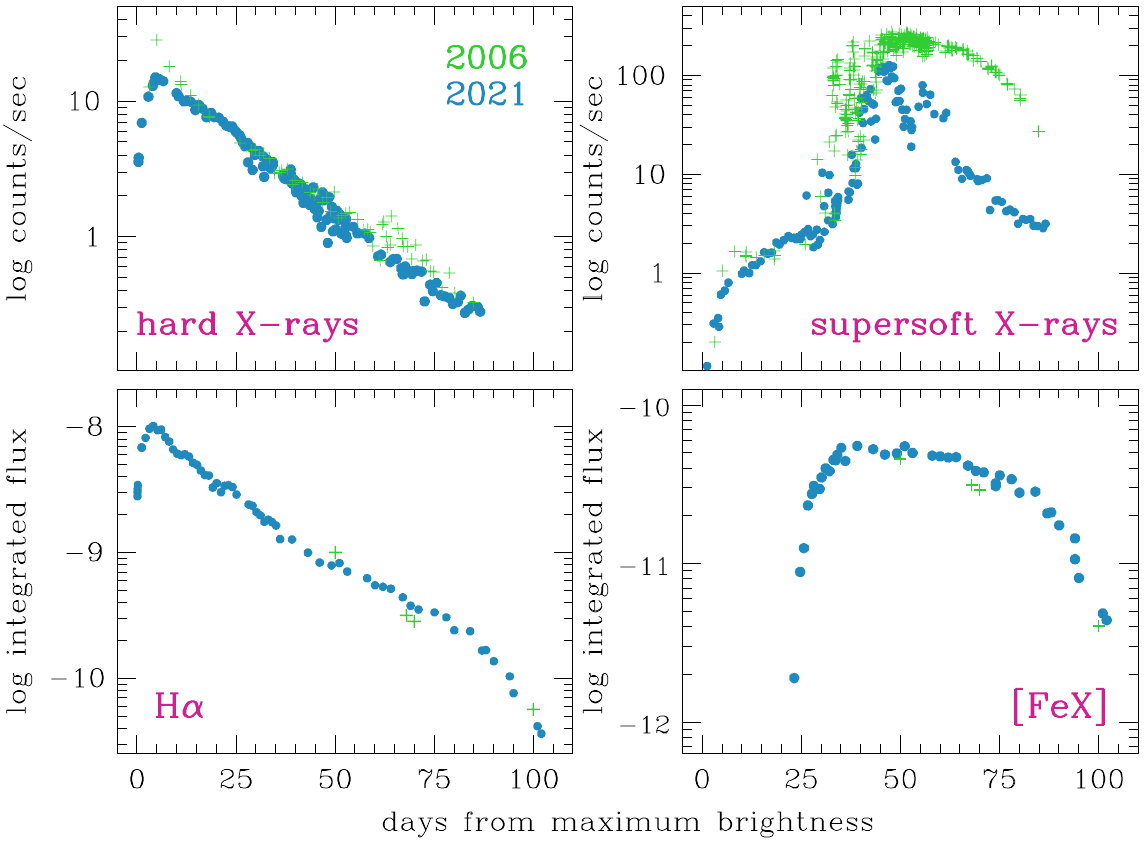}}
\caption{Comparing the evolution in flux for the 2006 and 2021 outbursts of
RS Oph for hard X-rays, super-soft X-rays, permitted and coronal emission
lines \citep[data from][]{2022MNRAS.514.1557P, 2022arXiv220301378M}. The
fluxes of optical emission lines are in erg\,cm$^{-2}$\,s$^{-1}$.}
\label{figKim}
\end{figure}

\subsection{X-rays}

The evolution in X-rays of RS~Oph has been monitored at $\sim$daily cadence
with the {\it Swift} satellite for both the 2006 and 2021 outbursts.  The
energy distributions of X-rays have been deconvolved into hard and supersoft
components by \citet{2022MNRAS.514.1557P}, from which we import the data
plotted on the top row of Fig.~\ref{figKim}.

\underline{\sc Supersoft X-rays}.  They come directly from the central WD
which is burning nuclearly at the surface, and become visible around
day +25 when the medium around the WD clears enough to turn optically thin. 
Coronal emission lines appear at the same time (cf. the flux evolution of
[FeX] 6375~\AA\ plotted in the lower-right panel of Fig.~\ref{figKim}),
because they form in the circumstellar material now exposed to the hard
radiation emanating from the burning WD.  The coronal emission lines behaved
very smoothly with time and very similar for the 2006 and 2021 eruptions,
supporting identical burning phases on the WD for the two outbursts.  The
sharp decline in flux of the coronal lines suggests that the nuclear burning ends
around day +85, which also coincides with the drop in supersoft X-rays
observed during the 2006 event.

The excitation of coronal lines integrates the WD output over the 4$\pi$
solid angle, and such lines are therefore well suited to trace the real
output of the burning WD.  On the contrary, supersoft X-rays - which are
seen coming directly from the burning WD - are particularly sensitive to
inhomogeneities in the absorbing material crossing the line-of-sight.  This
explains the rapid switch on/off episodes at the start of the SSS phase in
2006 \citep{2011ApJ...727..124O}.  On the contrary, the SSS started smoothly
in 2021 but significantly under-performed at later epochs compared to 2006;
such difference has been modeled by \citet{2023A&A...670A.131N} in terms of
different column density of absorbing material at the two outbursts.  The
opposite behavior of supersoft X-rays seen in 2006 and 2021, both during
their rise to maximum and decline from, probably relates to the fact that
the WD+RG binary was at opposite orbital phases at the time of the two
eruptions.  Therefore the line-of-sight to the WD may have probed at grazing
incidence, in one eruption the inner wall of the cavity shaped into DEOP by
the ejecta, and instead the outer atmosphere/inner wind of the RG at the
other.

\underline{\sc Hard X-rays}.  The hard X-rays originate from the ejecta
colliding with the pre-existing material.  The flux evolution of hard X-rays
is identical for 2006 and 2021 (upper-left panel of Fig.~\ref{figKim}),
suggesting that RG wind had similarly refilled the circumstellar space after
the previous eruption and that the ejecta impacted with the same mass and
velocity (momentum).  The decline in hard X-ray is continuous, very smooth,
and without glitches, indicating a similarly smooth, continuous and extended
distribution of the target slow material.  This is the same conclusion we
drawn above from the FWHM evolution of H$\alpha$, a fact suggesting that
H$\alpha$ and hard X-rays come from the same location within the binary
system.  This is also supported by the time-dependence of the {\em flux}
radiated in H$\alpha$, which is plotted in the lower-left panel of
Fig.~\ref{figKim}, which is exactly the same as that of hard X-rays from the
panel above.

\subsection{Summary: a multi-wavelength 3D picture}

\begin{figure}
\centerline{\includegraphics[width=\textwidth,clip=]{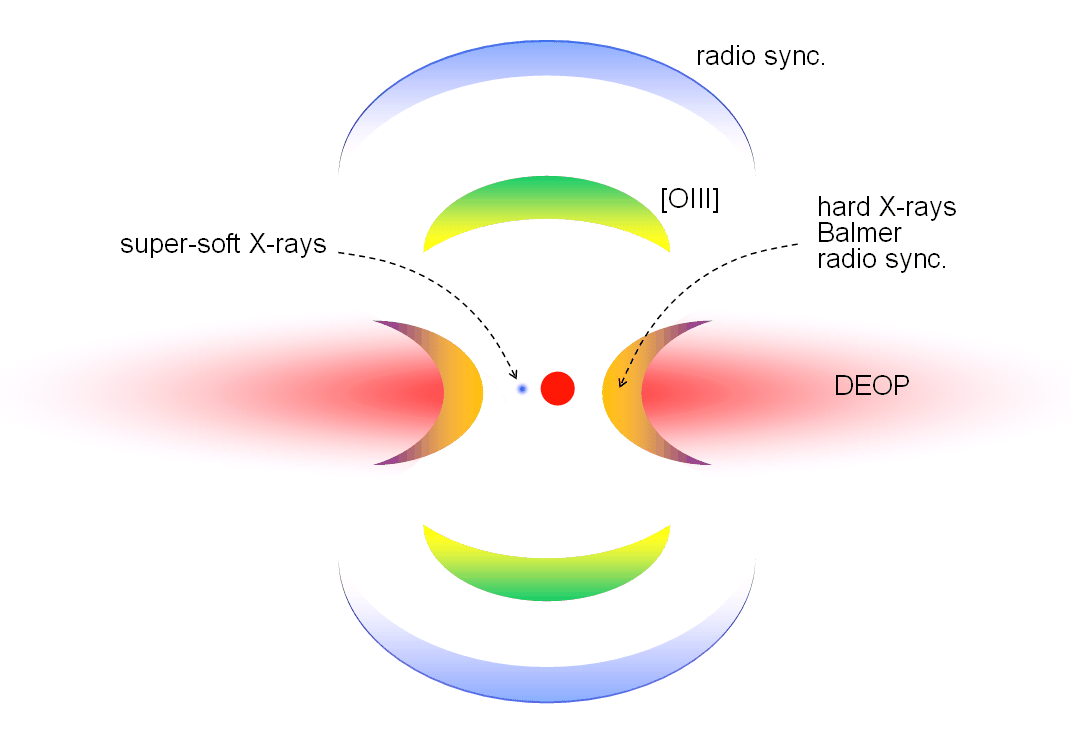}}
\caption{Artistic impression showing the location
of key emission features in RS~Oph around day +35 of the outburst
(drawing not to scale). The WD + RG binary is at the center, DEOP 
(density enhancement on the orbital plane) is seen edge-on, and the
polar directions run vertically.}
\label{figmodel}
\end{figure}

Putting all together the information from the previous sections about the
size, velocity and time evolution at radio, optical, and X-rays, a 3D
picture of RS~Oph (serving also as a guideline for symbiotic novae in
general) can be assembled as illustrated in Fig.~\ref{figmodel}.

The WD ejecta are lunched symmetrically by the WD at
10,000 to 12,000 km\,sec$^{-1}$ initial velocity.  They interact with the
pre-existing circumstellar material primarily at two distinct locations.

One location is very close to the central binary, at the inner radius of
DEOP (the density enhancement on the orbital plane formed by the
gravitational focusing of the RG wind by the orbiting WD).  The high density
and long radial extent of DEOP keeps the ejecta decelerating very smoothly for
weeks/months, and the shock interface travels only a short
distance, $\sim$15 AU in 100 days at which point the ejecta have been
practically arrested (velocities $\leq$150~km\,sec$^{-1}$).  The DEOP/ejecta
interface is the location from where originates most of the hard X-rays, the
central radio-synchrotron component, and the permitted optical emission
lines.  This interface is also probably the place where the high-energy
$\gamma$-rays (visible in the first days of the outburst) are produced.

The other location is toward the poles of the orbital plane, where the
density of pre-existing material is much lower compared to DEOP. 
Perpendicular to the orbital plane the deceleration is much less effective,
and the ejecta launched in that direction, after 100 days are still
traveling at thousands of km\,sec$^{-1}$, having crossed hundreds of AU.
The shocked outer edge of the resulting bipolar lobes
is the location of the synchrotron emission dominating at radio
cm-wavelengths. 

The forbidden emission lines are observed in RS Oph to move at constant and
high velocity, much larger that for permitted lines, as illustrated in the
lower-left panel of Fig.~\ref{figfwhm}.  With all probability they form in
the inner regions of the bipolar lobes, which move at a reduced velocity
compared with their outer edges, turning visible once the local density
dilutes below the critical value for collisional de-excitation of their upper
metastable levels.

The cavity in the pre-existing circumstellar material shaped by ejecta
plowing through them is efficiently and quickly refilled by the wind of the
red giant: each outburst of RS Oph has behaved exactly the same, even those
separated by just 9 years (like 1958 and 1967).  The accretion disk around
the WD is even quicker to reform, $\sim$250 days being enough to restore
both pre-outburst disk brightness and flickering activity
\citep{2022RNAAS...6..103M}.  

\section{An imminent eruption of T CrB ?}

The two known outbursts of T CrB in 1866 and 1946 are separated by 80 yrs,
and other 80yrs will bring us to 2026.  The 1946 eruption was preceded by a
SAP phase lasting $\sim$8 yrs, and a new SAP characterized T CrB from 2015
to 2023 \citep{2016NewA...47....7M} leading to the natural conclusion that a
new eruption may be imminent.  Such a view has been shared by many
\citep[eg.][]{2023MNRAS.524.3146S,2023A&A...680L..18Z}, with the beneficial
results of an intensified pre-eruption monitoring of T CrB and the
submission of proposals to most of the ground and space observing
facilities; proposals waiting to be triggered at the very first announcement
of the outburst !  Such an alertness from the community, the expected peak
$V$$\sim$3~mag brightness, and the proximity of T CrB to us (just 0.9 kpc
distance and negligible interstellar reddening), with all probabilities will
results is such a wealth of multi-wavelength information to keep people busy
for long in understanding and modeling them, leading to a significant leap
forward in our understanding of novae in general and their symbiotic
subclass in particoular.

\acknowledgements
We acknowledge the INAF 2023 Minigrant funding program and the Conference
Organizers for financial support.  We thank K.  Page, F.  Walter, R.  Lico,
and P.  Valisa for permission to use their data in drawing some of the
figures.
\bibliography{paper_rev}

\begin{thebibliography}{71}
\expandafter\ifx\csname natexlab\endcsname\relax\def\natexlab#1{#1}\fi

%ADS_ID 2010Sci...329..817A
\bibitem[{{Abdo} {et~al.}(2010){Abdo}, {Ackermann}, {Ajello}, {Atwood},
  {Baldini}, {Ballet}, {Barbiellini}, {Bastieri}, {Bechtol}, {Bellazzini},
  {Berenji}, {Blandford}, {Bloom}, {Bonamente}, {Borgland}, {Bouvier},
  {Brandt}, {Bregeon}, {Brez}, {Brigida}, {Bruel}, {Buehler}, {Burnett},
  {Buson}, {Caliandro}, {Cameron}, {Caraveo}, {Carrigan}, {Casandjian},
  {Cecchi}, {Celik}, {Charles}, {Chaty}, {Chekhtman}, {Cheung}, {Chiang},
  {Ciprini}, {Claus}, {Cohen-Tanugi}, {Conrad}, {Corbel}, {Corbet}, {DeCesar},
  {den Hartog}, {Dermer}, {de Palma}, {Digel}, {Donato}, {do Couto e Silva},
  {Drell}, {Dubois}, {Dubus}, {Dumora}, {Favuzzi}, {Fegan}, {Ferrara},
  {Fortin}, {Frailis}, {Fuhrmann}, {Fukazawa}, {Funk}, {Fusco}, {Gargano},
  {Gasparrini}, {Gehrels}, {Germani}, {Giglietto}, {Giordano}, {Giroletti},
  {Glanzman}, {Godfrey}, {Grenier}, {Grondin}, {Grove}, {Guiriec}, {Hadasch},
  {Harding}, {Hayashida}, {Hays}, {Healey}, {Hill}, {Horan}, {Hughes}, {Itoh},
  {Jean}, {J{\'o}hannesson}, {Johnson}, {Johnson}, {Johnson}, {Johnson},
  {Kamae}, {Katagiri}, {Kataoka}, {Kerr}, {Kn{\"o}dlseder}, {Koerding}, {Kuss},
  {Lande}, {Latronico}, {Lee}, {Lemoine-Goumard}, {Garde}, {Longo}, {Loparco},
  {Lott}, {Lovellette}, {Lubrano}, {Makeev}, {Mazziotta}, {McConville},
  {McEnery}, {Mehault}, {Michelson}, {Mizuno}, {Moiseev}, {Monte}, {Monzani},
  {Morselli}, {Moskalenko}, {Murgia}, {Nakamori}, {Naumann-Godo}, {Nestoras},
  {Nolan}, {Norris}, {Nuss}, {Ohno}, {Ohsugi}, {Okumura}, {Omodei}, {Orlando},
  {Ormes}, {Ozaki}, {Paneque}, {Panetta}, {Parent}, {Pelassa}, {Pepe},
  {Pesce-Rollins}, {Piron}, {Porter}, {Rain{\'o}}, {Rando}, {Ray}, {Razzano},
  {Razzaque}, {Rea}, {Reimer}, {Reimer}, {Reposeur}, {Ripken}, {Ritz},
  {Romani}, {Roth}, {Sadrozinski}, {Sander}, {Parkinson}, {Scargle},
  {Schinzel}, {Sgr{\`o}}, {Shaw}, {Siskind}, {Smith}, {Smith}, {Sokolovsky},
  {Spandre}, {Spinelli}, {Stawarz}, {Strickman}, {Suson}, {Takahashi},
  {Takahashi}, {Tanaka}, {Tanaka}, {Thayer}, {Thayer}, {Thompson}, {Torres},
  {Tosti}, {Tramacere}, {Uchiyama}, {Usher}, {Vandenbroucke}, {Vasileiou},
  {Vilchez}, {Vitale}, {Waite}, {Wallace}, {Wang}, {Winer}, {Wolff}, {Wood},
  {Yang}, {Ylinen}, {Ziegler}, {Maehara}, {Nishiyama}, {Kabashima}, {Bach},
  {Bower}, {Falcone}, {Forster}, {Henden}, {Kawabata}, {Koubsky}, {Mukai},
  {Nelson}, {Oates}, {Sakimoto}, {Sasada}, {Shenavrin}, {Shore}, {Skinner},
  {Sokoloski}, {Stroh}, {Tatarnikov}, {Uemura}, {Wahlgren}, {Yamanaka}, \&
  {Fermi LAT Collaboration}}]{2010Sci...329..817A}
{Abdo}, A.~A., {Ackermann}, M., {Ajello}, M., {et~al.}, {Gamma-Ray Emission
  Concurrent with the Nova in the Symbiotic Binary V407 Cygni}. 2010, {\it
  Science}, {\bf 329}, 817, DOI: 10.1126/science.1192537

%ADS_ID 2022NatAs...6..689A
\bibitem[{{Acciari} {et~al.}(2022){Acciari}, {Ansoldi}, {Antonelli}, {Arbet
  Engels}, {Artero}, {Asano}, {Baack}, {Babi{\'c}}, {Baquero}, {Barres de
  Almeida}, {Barrio}, {Batkovi{\'c}}, {Becerra Gonz{\'a}lez}, {Bednarek},
  {Bellizzi}, {Bernardini}, {Bernardos}, {Berti}, {Besenrieder},
  {Bhattacharyya}, {Bigongiari}, {Biland}, {Blanch}, {B{\"o}kenkamp},
  {Bonnoli}, {Bo{\v{s}}njak}, {Busetto}, {Carosi}, {Ceribella}, {Cerruti},
  {Chai}, {Chilingarian}, {Cikota}, {Colak}, {Colombo}, {Contreras}, {Cortina},
  {Covino}, {D'Amico}, {D'Elia}, {Da Vela}, {Dazzi}, {De Angelis}, {De Lotto},
  {Del Popolo}, {Delfino}, {Delgado}, {Delgado Mendez}, {Depaoli}, {Di Pierro},
  {Di Venere}, {Do Souto Espi{\~n}eira}, {Prester}, {Donini}, {Dorner}, {Doro},
  {Elsaesser}, {Fallah Ramazani}, {Fari{\~n}a Alonso}, {Fattorini}, {Fonseca},
  {Font}, {Fruck}, {Fukami}, {Fukazawa}, {Garc{\'\i}a L{\'o}pez},
  {Garczarczyk}, {Gasparyan}, {Gaug}, {Giglietto}, {Giordano}, {Gliwny},
  {Godinovi{\'c}}, {Green}, {Green}, {Hadasch}, {Hahn}, {Hassan}, {Heckmann},
  {Herrera}, {Hoang}, {Hrupec}, {H{\"u}tten}, {Inada}, {Ishio}, {Iwamura},
  {Jim{\'e}nez Mart{\'\i}nez}, {Jormanainen}, {Jouvin}, {Kerszberg},
  {Kobayashi}, {Kubo}, {Kushida}, {Lamastra}, {Lelas}, {Leone}, {Lindfors},
  {Linhoff}, {Lombardi}, {Longo}, {L{\'o}pez-Coto}, {L{\'o}pez-Moya},
  {L{\'o}pez-Oramas}, {Loporchio}, {Machado de Oliveira Fraga}, {Maggio},
  {Majumdar}, {Makariev}, {Mallamaci}, {Maneva}, {Manganaro}, {Mannheim},
  {Maraschi}, {Mariotti}, {Mart{\'\i}nez}, {Mas Aguilar}, {Mazin}, {Menchiari},
  {Mender}, {Mi{\'c}anovi{\'c}}, {Miceli}, {Miener}, {Miranda}, {Mirzoyan},
  {Molina}, {Moralejo}, {Morcuende}, {Moreno}, {Moretti}, {Nakamori}, {Nava},
  {Neustroev}, {Nievas Rosillo}, {Nigro}, {Nilsson}, {Nishijima}, {Noda},
  {Nozaki}, {Ohtani}, {Oka}, {Otero-Santos}, {Paiano}, {Palatiello}, {Paneque},
  {Paoletti}, {Paredes}, {Pavleti{\'c}}, {Pe{\~n}il}, {Persic}, {Pihet}, {Prada
  Moroni}, {Prandini}, {Priyadarshi}, {Puljak}, {Rhode}, {Rib{\'o}}, {Rico},
  {Righi}, {Rugliancich}, {Sahakyan}, {Saito}, {Sakurai}, {Satalecka},
  {Saturni}, {Schleicher}, {Schmidt}, {Schweizer}, {Sitarek},
  {{\v{S}}nidari{\'c}}, {Sobczynska}, {Spolon}, {Stamerra},
  {Stri{\v{s}}kovi{\'c}}, {Strom}, {Strzys}, {Suda}, {Suri{\'c}}, {Takahashi},
  {Takeishi}, {Tavecchio}, {Temnikov}, {Terzi{\'c}}, {Teshima}, {Tosti},
  {Truzzi}, {Tutone}, {Ubach}, {van Scherpenberg}, {Vanzo}, {Vazquez Acosta},
  {Ventura}, {Verguilov}, {Vigorito}, {Vitale}, {Vovk}, {Will}, {Wunderlich},
  {Yamamoto}, {Zari{\'c}}, {Ambrosino}, {Cecconi}, {Catanzaro}, {Ferrara},
  {Frasca}, {Munari}, {Giustolisi}, {Alonso-Santiago}, {Giarrusso}, {Munari},
  \& {Valisa}}]{2022NatAs...6..689A}
{Acciari}, V.~A., {Ansoldi}, S., {Antonelli}, L.~A., {et~al.}, {Proton
  acceleration in thermonuclear nova explosions revealed by gamma rays}. 2022,
  {\it Nature Astronomy}, {\bf 6}, 689, DOI: 10.1038/s41550-022-01640-z

%ADS_ID 1980MNRAS.192..521A
\bibitem[{{Allen}(1980)}]{1980MNRAS.192..521A}
{Allen}, D.~A., {On the late-type components of slow novae and symbiotic
  stars.} 1980, {\it \mnras}, {\bf 192}, 521, DOI: 10.1093/mnras/192.3.521

%ADS_ID 2024ApJ...963...17A
\bibitem[{{Arseneau} {et~al.}(2024){Arseneau}, {Chandra}, {Hwang}, {Zakamska},
  {Pallathadka}, {Crumpler}, {Hermes}, {El-Badry}, {Rix}, {Stassun},
  {G{\"a}nsicke}, {Brownstein}, \& {Morrison}}]{2024ApJ...963...17A}
{Arseneau}, S., {Chandra}, V., {Hwang}, H.-C., {et~al.}, {Measuring the
  Mass{\textendash}Radius Relation of White Dwarfs Using Wide Binaries}. 2024,
  {\it \apj}, {\bf 963}, 17, DOI: 10.3847/1538-4357/ad2168

%ADS_ID 2014ApJ...785L..11B
\bibitem[{{Banerjee} {et~al.}(2014){Banerjee}, {Joshi}, {Venkataraman},
  {Ashok}, {Marion}, {Hsiao}, \& {Raj}}]{2014ApJ...785L..11B}
{Banerjee}, D.~P.~K., {Joshi}, V., {Venkataraman}, V., {et~al.}, {Near-IR
  Studies of Recurrent Nova V745 Scorpii during its 2014 Outburst}. 2014, {\it
  \apjl}, {\bf 785}, L11, DOI: 10.1088/2041-8205/785/1/L11

%ADS_ID 1989clno.conf.....B
\bibitem[{{Bode} \& {Evans}(1989)}]{1989clno.conf.....B}
{Bode}, M.~F. \& {Evans}, A., eds. 1989, {\it {Classical novae}} (John Wiley \&
  Sons)

%ADS_ID 2008clno.book.....B
\bibitem[{{Bode} \& {Evans}(2008)}]{2008clno.book.....B}
{Bode}, M.~F. \& {Evans}, A., eds. 2008, {\it {Classical Novae}} (Cambridge
  Univ. Press)

%ADS_ID 2009A&A...497..815B
\bibitem[{{Brandi} {et~al.}(2009){Brandi}, {Quiroga}, {Miko{\l}ajewska},
  {Ferrer}, \& {Garc{\'\i}a}}]{2009A&A...497..815B}
{Brandi}, E., {Quiroga}, C., {Miko{\l}ajewska}, J., {Ferrer}, O.~E., \&
  {Garc{\'\i}a}, L.~G., {Spectroscopic orbits and variations of RS Ophiuchi}.
  2009, {\it \aap}, {\bf 497}, 815, DOI: 10.1051/0004-6361/200811417

%ADS_ID 2004ESASP1283.....C
\bibitem[{{Cassatella} {et~al.}(2004){Cassatella}, {Gonz{\'a}lez-Riestra}, \&
  {Selvelli}}]{2004ESASP1283.....C}
{Cassatella}, A., {Gonz{\'a}lez-Riestra}, R., \& {Selvelli}, P., eds. 2004, ESA
  Special Publication, Vol. {\bf  1283}, {\it {INES Access Guide No. 3 -
  Classical Novae}}

%ADS_ID 1990apcb.conf..373C
\bibitem[{{Cassatella} {et~al.}(1990){Cassatella}, {Selvelli}, {Gilmozzi},
  {Bianchini}, \& {Friedjung}}]{1990apcb.conf..373C}
{Cassatella}, A., {Selvelli}, P.~L., {Gilmozzi}, R., {Bianchini}, A., \&
  {Friedjung}, M., {IUE observations of faint old novae.} 1990, in {\it
  Accretion-Powered Compact Binaries}, ed. C.~W. {Mauche} (Cambridge Univ.
  Press), 373--376

%ADS_ID 1990LNP...369.....C
\bibitem[{{Cassatella} \& {Viotti}(1990)}]{1990LNP...369.....C}
{Cassatella}, A. \& {Viotti}, R., eds. 1990, Lecture Notes in Physics, Vol.
  {\bf  369}, {\it {Physics of Classical Novae}} (Springer-Verlag)

%ADS_ID 1997PASP..109..345D
\bibitem[{{Downes} {et~al.}(1997){Downes}, {Webbink}, \&
  {Shara}}]{1997PASP..109..345D}
{Downes}, R., {Webbink}, R.~F., \& {Shara}, M.~M., {A Catalog and Atlas of
  Cataclysmic Variables-Second Edition}. 1997, {\it \pasp}, {\bf 109}, 345,
  DOI: 10.1086/133900

%ADS_ID 1993PASP..105..127D
\bibitem[{{Downes} \& {Shara}(1993)}]{1993PASP..105..127D}
{Downes}, R.~A. \& {Shara}, M.~M., {A Catalog of Cataclysmic Variables}. 1993,
  {\it \pasp}, {\bf 105}, 127, DOI: 10.1086/133139

%ADS_ID 2001PASP..113..764D
\bibitem[{{Downes} {et~al.}(2001){Downes}, {Webbink}, {Shara}, {Ritter},
  {Kolb}, \& {Duerbeck}}]{2001PASP..113..764D}
{Downes}, R.~A., {Webbink}, R.~F., {Shara}, M.~M., {et~al.}, {A Catalog and
  Atlas of Cataclysmic Variables: The Living Edition}. 2001, {\it \pasp}, {\bf
  113}, 764, DOI: 10.1086/320802

%ADS_ID 1987rcag.book.....D
\bibitem[{{Duerbeck}(1987)}]{1987rcag.book.....D}
{Duerbeck}, H.~W. 1987, {\it {A reference catalogue and atlas of galactic
  novae}} (D. Reidel Publ.)

%ADS_ID 2008ASPC..401.....E
\bibitem[{{Evans} {et~al.}(2008){Evans}, {Bode}, {O'Brien}, \&
  {Darnley}}]{2008ASPC..401.....E}
{Evans}, A., {Bode}, M.~F., {O'Brien}, T.~J., \& {Darnley}, M.~J., eds. 2008,
  Astronomical Society of the Pacific Conference Series, Vol. {\bf  401}, {\it
  {RS Ophiuchi (2006) and the Recurrent Nova Phenomenon}}

%ADS_ID 2003ARA&A..41..517F
\bibitem[{{Ferland}(2003)}]{2003ARA&A..41..517F}
{Ferland}, G.~J., {Quantitative Spectroscopy of Photoionized Clouds}. 2003,
  {\it \araa}, {\bf 41}, 517, DOI: 10.1146/annurev.astro.41.011802.094836

%ADS_ID 1982ApJ...257..767F
\bibitem[{{Fujimoto}(1982{\natexlab{a}})}]{1982ApJ...257..767F}
{Fujimoto}, M.~Y., {A Theory of Hydrogen Shell Flashes on Accreting White
  Dwarfs - Part Two - the Stable Shell Burning and the Recurrence Period of
  Shell Flashes}. 1982{\natexlab{a}}, {\it \apj}, {\bf 257}, 767, DOI:
  10.1086/160030

%ADS_ID 1982ApJ...257..752F
\bibitem[{{Fujimoto}(1982{\natexlab{b}})}]{1982ApJ...257..752F}
{Fujimoto}, M.~Y., {A theory of hydrogen shell flashes on accreting white
  dwarfs. I - Their progress and the expansion of the envelope. II - The stable
  shell burning and the recurrence period of shell flashes}.
  1982{\natexlab{b}}, {\it \apj}, {\bf 257}, 752, DOI: 10.1086/160029

%ADS_ID 2016A&A...595A...1G
\bibitem[{{Gaia Collaboration}(2016)}]{2016A&A...595A...1G}
{Gaia Collaboration}, {The Gaia mission}. 2016, {\it \aap}, {\bf 595}, A1, DOI:
  10.1051/0004-6361/201629272

%ADS_ID 2023A&A...674A...1G
\bibitem[{{Gaia Collaboration}(2023)}]{2023A&A...674A...1G}
{Gaia Collaboration}, {Gaia Data Release 3. Summary of the content and survey
  properties}. 2023, {\it \aap}, {\bf 674}, A1, DOI:
  10.1051/0004-6361/202243940

%ADS_ID 1957gano.book.....G
\bibitem[{{Gaposchkin}(1957)}]{1957gano.book.....G}
{Gaposchkin}, C. H.~P. 1957, {\it {The Galactic Novae}} (Dover Pub.)

%ADS_ID 2020A&A...638A.130G
\bibitem[{{Giroletti} {et~al.}(2020){Giroletti}, {Munari}, {K{\"o}rding},
  {Mioduszewski}, {Sokoloski}, {Cheung}, {Corbel}, {Schinzel}, {Sokolovsky}, \&
  {O'Brien}}]{2020A&A...638A.130G}
{Giroletti}, M., {Munari}, U., {K{\"o}rding}, E., {et~al.}, {Very long baseline
  interferometry imaging of the advancing ejecta in the first gamma-ray nova
  V407 Cygni}. 2020, {\it \aap}, {\bf 638}, A130, DOI:
  10.1051/0004-6361/202038142

%ADS_ID 2022Sci...376...77H
\bibitem[{{H.~E.~S.~S. Collaboration} {et~al.}(2022){H.~E.~S.~S.
  Collaboration}, {Aharonian}, {Ait Benkhali}, {Ang{\"u}ner}, {Ashkar},
  {Backes}, {Baghmanyan}, {Barbosa Martins}, {Batzofin}, {Becherini}, {Berge},
  {Bernl{\"o}hr}, {Bi}, {B{\"o}ttcher}, {Boisson}, {Bolmont}, {de Bony de
  Lavergne}, {Breuhaus}, {Brose}, {Brun}, {Caroff}, {Casanova}, {Cerruti},
  {Chand}, {Chen}, {Cotter}, {Damascene Mbarubucyeye}, {Djannati-Ata{\"\i}},
  {Dmytriiev}, {Doroshenko}, {Duffy}, {Egberts}, {Ernenwein}, {Fegan},
  {Feijen}, {Fiasson}, {Fichet de Clairfontaine}, {Fontaine},
  {F{\"u}{\ss}ling}, {Funk}, {Gabici}, {Gallant}, {Ghafourizadeh}, {Giavitto},
  {Giunti}, {Glawion}, {Glicenstein}, {Grondin}, {Hermann}, {Hinton},
  {H{\"o}rbe}, {Hofmann}, {Hoischen}, {Holch}, {Holler}, {Horns}, {Huang},
  {Jamrozy}, {Jankowsky}, {Jung-Richardt}, {Kasai}, {Katarzy{\'n}ski}, {Katz},
  {Khangulyan}, {Kh{\'e}lifi}, {Klepser}, {Klu{\'z}niak}, {Komin}, {Konno},
  {Kosack}, {Kostunin}, {Le Stum}, {Lemi{\`e}re}, {Lemoine-Goumard}, {Lenain},
  {Leuschner}, {Lohse}, {Luashvili}, {Lypova}, {Mackey}, {Malyshev},
  {Malyshev}, {Marandon}, {Marchegiani}, {Marcowith}, {Mart{\'\i}-Devesa},
  {Marx}, {Maurin}, {Meyer}, {Mitchell}, {Moderski}, {Mohrmann}, {Montanari},
  {Moulin}, {Muller}, {Murach}, {Nakashima}, {de Naurois}, {Nayerhoda},
  {Niemiec}, {Priyana Noel}, {O{\textquoteright}Brien}, {Ohm}, {Olivera-Nieto},
  {de Ona Wilhelmi}, {Ostrowski}, {Panny}, {Panter}, {Parsons}, {Peron},
  {Pita}, {Poireau}, {Prokhorov}, {Prokoph}, {P{\"u}hlhofer}, {Punch},
  {Quirrenbach}, {Reichherzer}, {Reimer}, {Reimer}, {Renaud}, {Reville},
  {Rieger}, {Rowell}, {Rudak}, {Rueda Ricarte}, {Ruiz-Velasco}, {Sahakian},
  {Sailer}, {Salzmann}, {Sanchez}, {Santangelo}, {Sasaki}, {Sch{\"a}fer},
  {Sch{\"u}ssler}, {Schutte}, {Schwanke}, {Senniappan}, {Shapopi}, {Simoni},
  {Sinha}, {Sol}, {Specovius}, {Spencer}, {Stawarz}, {Steinmassl}, {Steppa},
  {Takahashi}, {Tanaka}, {Taylor}, {Terrier}, {Thorpe-Morgan}, {Tsirou},
  {Tsuji}, {Tuffs}, {Uchiyama}, {Unbehaun}, {van Eldik}, {van Soelen}, {Veh},
  {Venter}, {Vink}, {Wagner}, {Werner}, {White}, {Wierzcholska}, {Wong},
  {Yusafzai}, {Zacharias}, {Zargaryan}, {Zdziarski}, {Zech}, {Zhu}, {Zouari},
  \& {{\.Z}ywucka}}]{2022Sci...376...77H}
{H.~E.~S.~S. Collaboration}, {Aharonian}, F., {Ait Benkhali}, F., {et~al.},
  {Time-resolved hadronic particle acceleration in the recurrent nova RS
  Ophiuchi}. 2022, {\it Science}, {\bf 376}, 77, DOI: 10.1126/science.abn0567

%ADS_ID 2002AIPC..637.....H
\bibitem[{{Hernanz} \& {Jos{\'e}}(2002)}]{2002AIPC..637.....H}
{Hernanz}, M. \& {Jos{\'e}}, J., eds. 2002, American Institute of Physics
  Conference Series, Vol. {\bf  637}, {\it {Classical Nova Explosions}} (AIP)

%ADS_ID 2013ApJ...770...28H
\bibitem[{{Hinkle} {et~al.}(2013){Hinkle}, {Fekel}, {Joyce}, \&
  {Wood}}]{2013ApJ...770...28H}
{Hinkle}, K.~H., {Fekel}, F.~C., {Joyce}, R.~R., \& {Wood}, P., {Infrared
  Spectroscopy of Symbiotic Stars. IX. D-type Symbiotic Novae}. 2013, {\it
  \apj}, {\bf 770}, 28, DOI: 10.1088/0004-637X/770/1/28

%ADS_ID 1982ApJ...259..244I
\bibitem[{{Iben}(1982)}]{1982ApJ...259..244I}
{Iben}, I., J., {Hot accreting white dwarfs in the quasi-static approximation}.
  1982, {\it \apj}, {\bf 259}, 244, DOI: 10.1086/160164

%ADS_ID 1986syst.book.....K
\bibitem[{{Kenyon}(1986)}]{1986syst.book.....K}
{Kenyon}, S.~J. 1986, {\it {The symbiotic stars}} (Cambridge Univ. Press)

%ADS_ID 1983ApJ...273..280K
\bibitem[{{Kenyon} \& {Truran}(1983)}]{1983ApJ...273..280K}
{Kenyon}, S.~J. \& {Truran}, J.~W., {The outbursts of symbiotic novae.} 1983,
  {\it \apj}, {\bf 273}, 280, DOI: 10.1086/161367

%ADS_ID 2003ARep...47..777K
\bibitem[{{Kolotilov} {et~al.}(2003){Kolotilov}, {Shenavrin}, {Shugarov}, \&
  {Yudin}}]{2003ARep...47..777K}
{Kolotilov}, E.~A., {Shenavrin}, V.~I., {Shugarov}, S.~Y., \& {Yudin}, B.~F.,
  {UBVJHKLM photometry of the symbiotic Mira V407 Cyg in 1998 2002}. 2003, {\it
  Astronomy Reports}, {\bf 47}, 777, DOI: 10.1134/1.1611218

%ADS_ID 2024A&A...692A.107L
\bibitem[{{Lico} {et~al.}(2024){Lico}, {Giroletti}, {Munari}, {O'Brien},
  {Marcote}, {Williams}, {Yang}, {Veres}, \& {Woudt}}]{2024A&A...692A.107L}
{Lico}, R., {Giroletti}, M., {Munari}, U., {et~al.}, {High-resolution imaging
  of the evolving bipolar outflows in symbiotic novae: The case of the RS
  Ophiuchi 2021 nova outburst}. 2024, {\it \aap}, {\bf 692}, A107, DOI:
  10.1051/0004-6361/202451364

%ADS_ID 2012ApJ...761...34L
\bibitem[{{Liimets} {et~al.}(2012){Liimets}, {Corradi},
  {Santander-Garc{\'\i}a}, {Villaver}, {Rodr{\'\i}guez-Gil}, {Verro}, \&
  {Kolka}}]{2012ApJ...761...34L}
{Liimets}, T., {Corradi}, R.~L.~M., {Santander-Garc{\'\i}a}, M., {et~al.}, {A
  Three-dimensional View of the Remnant of Nova Persei 1901 (GK Per)}. 2012,
  {\it \apj}, {\bf 761}, 34, DOI: 10.1088/0004-637X/761/1/34

%ADS_ID 2012BaltA..21...88M
\bibitem[{{Mohamed} \& {Podsiadlowski}(2012)}]{2012BaltA..21...88M}
{Mohamed}, S. \& {Podsiadlowski}, P., {Mass Transfer in Mira-type Binaries}.
  2012, {\it Baltic Astronomy}, {\bf 21}, 88, DOI: 10.1515/astro-2017-0362

%ADS_ID 2024MNRAS.534.1227M
\bibitem[{{Molina} {et~al.}(2024){Molina}, {Chomiuk}, {Linford}, {Aydi},
  {Mioduszewski}, {Mukai}, {Sokolovsky}, {Strader}, {Craig}, {Dong}, {Harris},
  {Nyamai}, {Rupen}, {Sokoloski}, {Walter}, {Weston}, \&
  {Williams}}]{2024MNRAS.534.1227M}
{Molina}, I., {Chomiuk}, L., {Linford}, J.~D., {et~al.}, {The symbiotic
  recurrent nova V745 Sco at radio wavelengths}. 2024, {\it \mnras}, {\bf 534},
  1227, DOI: 10.1093/mnras/stae2093

%ADS_ID 2022ApJ...926..100M
\bibitem[{{Montez} {et~al.}(2022){Montez}, {Luna}, {Mukai}, {Sokoloski}, \&
  {Kastner}}]{2022ApJ...926..100M}
{Montez}, R., {Luna}, G.~J.~M., {Mukai}, K., {Sokoloski}, J.~L., \& {Kastner},
  J.~H., {Expanding Bipolar X-Ray Structure After the 2006 Eruption of RS Oph}.
  2022, {\it \apj}, {\bf 926}, 100, DOI: 10.3847/1538-4357/ac4583

%ADS_ID 2019arXiv190901389M
\bibitem[{{Munari}(2019)}]{2019arXiv190901389M}
{Munari}, U., {The Symbiotic Stars}. 2019, in {\it The Impact of Binary Stars
  on Stellar Evolution}, ed. G.~{Beccari} \& H.~{Boffin}, arXiv:1909.01389

%ADS_ID 2016NewA...47....7M
\bibitem[{{Munari} {et~al.}(2016){Munari}, {Dallaporta}, \&
  {Cherini}}]{2016NewA...47....7M}
{Munari}, U., {Dallaporta}, S., \& {Cherini}, G., {The 2015 super-active state
  of recurrent nova T CrB and the long term evolution after the 1946 outburst}.
  2016, {\it \na}, {\bf 47}, 7, DOI: 10.1016/j.newast.2016.01.002

%ADS_ID 2022A&A...666L...6M
\bibitem[{{Munari} {et~al.}(2022){Munari}, {Giroletti}, {Marcote}, {O'Brien},
  {Veres}, {Yang}, {Williams}, \& {Woudt}}]{2022A&A...666L...6M}
{Munari}, U., {Giroletti}, M., {Marcote}, B., {et~al.}, {Radio interferometric
  imaging of RS Oph bipolar ejecta for the 2021 nova outburst}. 2022, {\it
  \aap}, {\bf 666}, L6, DOI: 10.1051/0004-6361/202244821

%ADS_ID 2011MNRAS.410L..52M
\bibitem[{{Munari} {et~al.}(2011){Munari}, {Joshi}, {Ashok}, {Banerjee},
  {Valisa}, {Milani}, {Siviero}, {Dallaporta}, \&
  {Castellani}}]{2011MNRAS.410L..52M}
{Munari}, U., {Joshi}, V.~H., {Ashok}, N.~M., {et~al.}, {The 2010 nova outburst
  of the symbiotic Mira V407 Cyg}. 2011, {\it \mnras}, {\bf 410}, L52, DOI:
  10.1111/j.1745-3933.2010.00979.x

%ADS_ID 2022RNAAS...6..103M
\bibitem[{{Munari} \& {Tabacco}(2022)}]{2022RNAAS...6..103M}
{Munari}, U. \& {Tabacco}, F., {Flickering Returns as RS Oph Reestablishes
  Quiescent Conditions Following its 2021 Nova Outburst}. 2022, {\it Research
  Notes of the American Astronomical Society}, {\bf 6}, 103, DOI:
  10.3847/2515-5172/ac72ae

%ADS_ID 2021MNRAS.505.6121M
\bibitem[{{Munari} {et~al.}(2021){Munari}, {Traven}, {Masetti}, {Valisa},
  {Righetti}, {Hambsch}, {Frigo}, {{\v{C}}otar}, {De Silva}, {Freeman},
  {Lewis}, {Martell}, {Sharma}, {Simpson}, {Ting}, {Wittenmyer}, \&
  {Zucker}}]{2021MNRAS.505.6121M}
{Munari}, U., {Traven}, G., {Masetti}, N., {et~al.}, {The GALAH survey and
  symbiotic stars - I. Discovery and follow-up of 33 candidate accreting-only
  systems}. 2021, {\it \mnras}, {\bf 505}, 6121, DOI: 10.1093/mnras/stab1620

%ADS_ID 2021arXiv210901101M
\bibitem[{{Munari} \& {Valisa}(2021)}]{2021arXiv210901101M}
{Munari}, U. \& {Valisa}, P., {The 2021 outburst of RS Oph. A pictorial atlas
  of the spectroscopic evolution: the first 18 days}. 2021, {\it arXiv
  e-prints}, arXiv:2109.01101, DOI: 10.48550/arXiv.2109.01101

%ADS_ID 2022arXiv220301378M
\bibitem[{{Munari} \& {Valisa}(2022)}]{2022arXiv220301378M}
{Munari}, U. \& {Valisa}, P., {The 2021 outburst of RS Oph: a pictorial atlas
  of the spectroscopic evolution. II. From day 19 to 102 (solar conjunction)}.
  2022, {\it arXiv e-prints}, arXiv:2203.01378, DOI: 10.48550/arXiv.2203.01378

%ADS_ID 2023A&A...670A.131N
\bibitem[{{Ness} {et~al.}(2023){Ness}, {Beardmore}, {Bode}, {Darnley},
  {Dobrotka}, {Drake}, {Magdolen}, {Munari}, {Osborne}, {Orio}, {Page}, \&
  {Starrfield}}]{2023A&A...670A.131N}
{Ness}, J.~U., {Beardmore}, A.~P., {Bode}, M.~F., {et~al.}, {High-resolution
  X-ray spectra of RS Ophiuchi (2006 and 2021): Revealing the cause of SSS
  variability}. 2023, {\it \aap}, {\bf 670}, A131, DOI:
  10.1051/0004-6361/202245269

%ADS_ID 2023MNRAS.523.1661N
\bibitem[{{Nyamai} {et~al.}(2023){Nyamai}, {Linford}, {Allison}, {Chomiuk},
  {Woudt}, {Ribeiro}, \& {Sarbadhicary}}]{2023MNRAS.523.1661N}
{Nyamai}, M.~M., {Linford}, J.~D., {Allison}, J.~R., {et~al.}, {Synchrotron
  emission from double-peaked radio light curves of the symbiotic recurrent
  nova V3890 Sgr}. 2023, {\it \mnras}, {\bf 523}, 1661, DOI:
  10.1093/mnras/stad1534

%ADS_ID 2008clno.book..285O
\bibitem[{{O'Brien} \& {Bode}(2008)}]{2008clno.book..285O}
{O'Brien}, T. \& {Bode}, M., {Resolved nebular remnants}. 2008, in {\it
  Classical Novae}, ed. M.~F. {Bode} \& A.~{Evans} (Cabridge Univ. Press),
  285--305

%ADS_ID 2011ApJ...727..124O
\bibitem[{{Osborne} {et~al.}(2011){Osborne}, {Page}, {Beardmore}, {Bode},
  {Goad}, {O'Brien}, {Starrfield}, {Rauch}, {Ness}, {Krautter}, {Schwarz},
  {Burrows}, {Gehrels}, {Drake}, {Evans}, \& {Eyres}}]{2011ApJ...727..124O}
{Osborne}, J.~P., {Page}, K.~L., {Beardmore}, A.~P., {et~al.}, {The Supersoft
  X-ray Phase of Nova RS Ophiuchi 2006}. 2011, {\it \apj}, {\bf 727}, 124, DOI:
  10.1088/0004-637X/727/2/124

%ADS_ID 2022MNRAS.514.1557P
\bibitem[{{Page} {et~al.}(2022){Page}, {Beardmore}, {Osborne}, {Munari},
  {Ness}, {Evans}, {Bode}, {Darnley}, {Drake}, {Kuin}, {O'Brien}, {Orio},
  {Shore}, {Starrfield}, \& {Woodward}}]{2022MNRAS.514.1557P}
{Page}, K.~L., {Beardmore}, A.~P., {Osborne}, J.~P., {et~al.}, {The 2021
  outburst of the recurrent nova RS Ophiuchi observed in X-rays by the Neil
  Gehrels Swift Observatory: a comparative study}. 2022, {\it \mnras}, {\bf
  514}, 1557, DOI: 10.1093/mnras/stac1295

%ADS_ID 1946ApJ...104...75P
\bibitem[{{Payne-Gaposchkin} \& {Wright}(1946)}]{1946ApJ...104...75P}
{Payne-Gaposchkin}, C. \& {Wright}, F.~W., {The Photographic Light-Curve of T
  Coronae Borealis.} 1946, {\it \apj}, {\bf 104}, 75, DOI: 10.1086/144834

%ADS_ID 2009ApJ...703.1955R
\bibitem[{{Ribeiro} {et~al.}(2009){Ribeiro}, {Bode}, {Darnley}, {Harman},
  {Newsam}, {O'Brien}, {Bohigas}, {Echevarr{\'\i}a}, \&
  {Bond}}]{2009ApJ...703.1955R}
{Ribeiro}, V.~A.~R.~M., {Bode}, M.~F., {Darnley}, M.~J., {et~al.}, {The
  Expanding Nebular Remnant of the Recurrent Nova RS Ophiuchi (2006). II.
  Modeling of Combined Hubble Space Telescope Imaging and Ground-based
  Spectroscopy}. 2009, {\it \apj}, {\bf 703}, 1955, DOI:
  10.1088/0004-637X/703/2/1955

%ADS_ID 2012clno.book.....S
\bibitem[{{Saikia} \& {Anupama}(2012)}]{2012clno.book.....S}
{Saikia}, D.~F. \& {Anupama}, G.~C., eds. 2012, Bull. Astr. Soc. India, Vol.
  {\bf ~40}, {\it {Novae from radio to gamma rays}} (Astron. Soc. India)

%ADS_ID 2023MNRAS.524.3146S
\bibitem[{{Schaefer}(2023)}]{2023MNRAS.524.3146S}
{Schaefer}, B.~E., {The B \& V light curves for recurrent nova T CrB from
  1842-2022, the unique pre- and post-eruption high-states, the complex period
  changes, and the upcoming eruption in 2025.5 {\ensuremath{\pm}} 1.3}. 2023,
  {\it \mnras}, {\bf 524}, 3146, DOI: 10.1093/mnras/stad735

%ADS_ID 2011ApJS..197...31S
\bibitem[{{Schwarz} {et~al.}(2011){Schwarz}, {Ness}, {Osborne}, {Page},
  {Evans}, {Beardmore}, {Walter}, {Helton}, {Woodward}, {Bode}, {Starrfield},
  \& {Drake}}]{2011ApJS..197...31S}
{Schwarz}, G.~J., {Ness}, J.-U., {Osborne}, J.~P., {et~al.}, {Swift X-Ray
  Observations of Classical Novae. II. The Super Soft Source Sample}. 2011,
  {\it \apjs}, {\bf 197}, 31, DOI: 10.1088/0067-0049/197/2/31

%ADS_ID 1993ApJ...410..260S
\bibitem[{{Seaquist} {et~al.}(1993){Seaquist}, {Krogulec}, \&
  {Taylor}}]{1993ApJ...410..260S}
{Seaquist}, E.~R., {Krogulec}, M., \& {Taylor}, A.~R., {A Highly Sensitive
  Radio Survey of Symbiotic Stars at 3.6 Centimeters}. 1993, {\it \apj}, {\bf
  410}, 260, DOI: 10.1086/172742

%ADS_ID 1990ApJ...349..313S
\bibitem[{{Seaquist} \& {Taylor}(1990)}]{1990ApJ...349..313S}
{Seaquist}, E.~R. \& {Taylor}, A.~R., {The Collective Radio Properties of
  Symbiotic Stars}. 1990, {\it \apj}, {\bf 349}, 313, DOI: 10.1086/168315

%ADS_ID 1984ApJ...284..202S
\bibitem[{{Seaquist} {et~al.}(1984){Seaquist}, {Taylor}, \&
  {Button}}]{1984ApJ...284..202S}
{Seaquist}, E.~R., {Taylor}, A.~R., \& {Button}, S., {A Radio Survey of
  Symbiotic Stars}. 1984, {\it \apj}, {\bf 284}, 202, DOI: 10.1086/162399

%ADS_ID 2011A&A...527A..98S
\bibitem[{{Shore} {et~al.}(2011){Shore}, {Wahlgren}, {Augusteijn}, {Liimets},
  {Page}, {Osborne}, {Beardmore}, {Koubsky}, {{\v{S}}lechta}, \&
  {Votruba}}]{2011A&A...527A..98S}
{Shore}, S.~N., {Wahlgren}, G.~M., {Augusteijn}, T., {et~al.}, {The
  spectroscopic evolution of the symbiotic-like recurrent nova V407 Cygni
  during its 2010 outburst. I. The shock and its evolution}. 2011, {\it \aap},
  {\bf 527}, A98, DOI: 10.1051/0004-6361/201015901

%ADS_ID 2015A&A...573A...8S
\bibitem[{{Skopal} \& {Carikov{\'a}}(2015)}]{2015A&A...573A...8S}
{Skopal}, A. \& {Carikov{\'a}}, Z., {Wind mass transfer in S-type symbiotic
  binaries. I. Focusing by the wind compression model}. 2015, {\it \aap}, {\bf
  573}, A8, DOI: 10.1051/0004-6361/201424779

%ADS_ID 2020A&A...636A..77S
\bibitem[{{Skopal} {et~al.}(2020){Skopal}, {Shugarov}, {Munari}, {Masetti},
  {Marchesini}, {Kom{\v{z}}{\'\i}k}, {Kundra}, {Shagatova}, {Tarasova}, {Buil},
  {Boussin}, {Shenavrin}, {Hambsch}, {Dallaporta}, {Frigo}, {Garde},
  {Zubareva}, {Dubovsk{\'y}}, \& {Kroll}}]{2020A&A...636A..77S}
{Skopal}, A., {Shugarov}, S.~Y., {Munari}, U., {et~al.}, {The path to Z
  And-type outbursts: The case of V426 Sagittae (HBHA 1704-05)}. 2020, {\it
  \aap}, {\bf 636}, A77, DOI: 10.1051/0004-6361/201937199

%ADS_ID 2006AJ....131.1163S
\bibitem[{{Skrutskie} {et~al.}(2006){Skrutskie}, {Cutri}, {Stiening},
  {Weinberg}, {Schneider}, {Carpenter}, {Beichman}, {Capps}, {Chester},
  {Elias}, {Huchra}, {Liebert}, {Lonsdale}, {Monet}, {Price}, {Seitzer},
  {Jarrett}, {Kirkpatrick}, {Gizis}, {Howard}, {Evans}, {Fowler}, {Fullmer},
  {Hurt}, {Light}, {Kopan}, {Marsh}, {McCallon}, {Tam}, {Van Dyk}, \&
  {Wheelock}}]{2006AJ....131.1163S}
{Skrutskie}, M.~F., {Cutri}, R.~M., {Stiening}, R., {et~al.}, {The Two Micron
  All Sky Survey (2MASS)}. 2006, {\it \aj}, {\bf 131}, 1163, DOI:
  10.1086/498708

%ADS_ID 2008ApJ...685L.137S
\bibitem[{{Sokoloski} {et~al.}(2008){Sokoloski}, {Rupen}, \&
  {Mioduszewski}}]{2008ApJ...685L.137S}
{Sokoloski}, J.~L., {Rupen}, M.~P., \& {Mioduszewski}, A.~J., {Uncovering the
  Nature of Nova Jets: A Radio Image of Highly Collimated Outflows from RS
  Ophiuchi}. 2008, {\it \apjl}, {\bf 685}, L137, DOI: 10.1086/592602

%ADS_ID 2021ATel14886....1S
\bibitem[{{Sokolovsky} {et~al.}(2021){Sokolovsky}, {Aydi}, {Chomiuk}, {Kawash},
  {Strader}, {Babul}, {Sokoloski}, {Mioduszewski}, {Linford}, {Mukai}, {Li},
  {O'Brien}, \& {Rupen}}]{2021ATel14886....1S}
{Sokolovsky}, K., {Aydi}, E., {Chomiuk}, L., {et~al.}, {VLA observations of the
  2021 eruption of RS Oph}. 2021, {\it The Astronomer's Telegram}, {\bf 14886},
  1

%ADS_ID 1989clno.conf...39S
\bibitem[{{Starrfield}(1989)}]{1989clno.conf...39S}
{Starrfield}, S., {Thermonuclear processes and the classical nova outburst.}
  1989, in {\it Classical Novae}, ed. M.~F. {Bode} \& A.~{Evans} (John Wiley \&
  Sons), 39--60

%ADS_ID 2008clno.book...77S
\bibitem[{{Starrfield} {et~al.}(2008){Starrfield}, {Iliadis}, \&
  {Hix}}]{2008clno.book...77S}
{Starrfield}, S., {Iliadis}, C., \& {Hix}, W.~R., {Thermonuclear processes}.
  2008, in {\it Classical Novae}, ed. M.~F. {Bode} \& A.~{Evans} (Cambridge
  Univ. Press), 77--101

%ADS_ID 2003ARep...47..889T
\bibitem[{{Tatarnikova} {et~al.}(2003){Tatarnikova}, {Marrese}, {Munari},
  {Tomov}, \& {Yudin}}]{2003ARep...47..889T}
{Tatarnikova}, A.~A., {Marrese}, P.~M., {Munari}, U., {Tomov}, T., \& {Yudin},
  B.~F., {Spectral observations of the symbiotic Mira variable V407 Cyg in 1993
  2002}. 2003, {\it Astronomy Reports}, {\bf 47}, 889, DOI: 10.1134/1.1626192

%ADS_ID 1984ApJ...286..263T
\bibitem[{{Taylor} \& {Seaquist}(1984)}]{1984ApJ...286..263T}
{Taylor}, A.~R. \& {Seaquist}, E.~R., {Radio emission from symbiotic stars : a
  binary model.} 1984, {\it \apj}, {\bf 286}, 263, DOI: 10.1086/162594

%ADS_ID 2016MNRAS.462.4435T
\bibitem[{{Tomov} {et~al.}(2016){Tomov}, {Stoyanov}, \&
  {Zamanov}}]{2016MNRAS.462.4435T}
{Tomov}, T.~V., {Stoyanov}, K.~A., \& {Zamanov}, R.~K., {AG Pegasi - now a
  classical symbiotic star in outburst?} 2016, {\it \mnras}, {\bf 462}, 4435,
  DOI: 10.1093/mnras/stw2012

%ADS_ID 1995cvs..book.....W
\bibitem[{{Warner}(1995)}]{1995cvs..book.....W}
{Warner}, B. {\it {Cataclysmic variable stars}}, , Vol. {\bf ~28} (Cambridge
  Univ. Press)

%ADS_ID 2021ATel14849....1W
\bibitem[{{Williams} {et~al.}(2021){Williams}, {O'Brien}, {Woudt}, {Nyamai},
  {Green}, {Titterington}, {Fender}, \& {Sivakoff}}]{2021ATel14849....1W}
{Williams}, D., {O'Brien}, T., {Woudt}, P., {et~al.}, {AMI-LA, e-MERLIN and
  MeerKAT radio detections of RS Oph in outburst}. 2021, {\it The Astronomer's
  Telegram}, {\bf 14849}, 1

%ADS_ID 2014ASPC..490.....W
\bibitem[{{Woudt} \& {Ribeiro}(2014)}]{2014ASPC..490.....W}
{Woudt}, P.~A. \& {Ribeiro}, V.~A.~R.~M., eds. 2014, Astronomical Society of
  the Pacific Conference Series, Vol. {\bf  490}, {\it {Stella Novae: Past and
  Future Decades}}

%ADS_ID 2023A&A...680L..18Z
\bibitem[{{Zamanov} {et~al.}(2023){Zamanov}, {Boeva}, {Latev}, {Semkov},
  {Minev}, {Kostov}, {Bode}, {Marchev}, \& {Marchev}}]{2023A&A...680L..18Z}
{Zamanov}, R., {Boeva}, S., {Latev}, G.~Y., {et~al.}, {Accretion in the
  recurrent nova T CrB: Linking the superactive state to the predicted
  outburst}. 2023, {\it \aap}, {\bf 680}, L18, DOI: 10.1051/0004-6361/202348372

\end{thebibliography}
\end{document}